\documentclass[english]{IEEEtran}
\usepackage[T1]{fontenc}
\usepackage[latin9]{inputenc}
\usepackage{amsthm}
\usepackage{amsmath}
\usepackage{amssymb}
\usepackage{esint}

\makeatletter
\theoremstyle{plain}
\newtheorem{thm}{\protect\theoremname}
\theoremstyle{plain}
\newtheorem{cor}[thm]{\protect\corollaryname}
\theoremstyle{remark}
\newtheorem{rem}[thm]{\protect\remarkname}
\theoremstyle{plain}
\newtheorem{lem}[thm]{\protect\lemmaname}

\usepackage{cite}
\allowdisplaybreaks
\makeatother

\usepackage{babel}
\providecommand{\corollaryname}{Corollary}
\providecommand{\lemmaname}{Lemma}
\providecommand{\remarkname}{Remark}
\providecommand{\theoremname}{Theorem}

\begin{document}

\title{Connectivity of Large Wireless Networks under A Generic Connection
Model%
\thanks{Some results in Section \ref{sec:connectivity necessary condition}
of this paper appeared in INFOCOM 2011 \cite{Mao11On}. Substantial
improvements have been made on the theoretical analysis in \cite{Mao11On}.%
}}

\author{Guoqiang Mao, \emph{Senior Member}%
\thanks{G. Mao is with the University of Sydney and National ICT Australia
 (email: guoqiang.mao@sydney.edu.au).%
}\emph{, IEEE} and Brian D.O. Anderson%
\thanks{B. D. O. Anderson is with the Australian National University and National
ICT Australia (email: brian.anderson@anu.edu.au).%
}, \emph{Life Fellow, IEEE}%
\thanks{This research is funded by ARC Discovery projects: DP110100538 and
DP120102030.%
}\emph{}%
\thanks{Copyright (c) 2012 IEEE. Personal use of this material is permitted.
However, permission to use this material for any other purposes must
be obtained from the IEEE by sending a request to pubs-permissions@ieee.org.%
}}
\maketitle
\begin{abstract}
This paper studies networks where all nodes are distributed on a unit
square $A\triangleq[-\frac{1}{2},\frac{1}{2}]^{2}$ following a Poisson
distribution with known density $\rho$ and a pair of nodes separated
by an Euclidean distance $x$ are directly connected with probability
$g_{r_{\rho}}(x)\triangleq g(x/r_{\rho})$, independent of the event
that any other pair of nodes are directly connected. Here $g:[0,\infty)\rightarrow[0,1]$
satisfies the conditions of rotational invariance, non-increasing
monotonicity, integral boundedness and $g\left(x\right)=o(1/(x^{2}\log^{2}x))$;
further, $r_{\rho}=\sqrt{(\log\rho+b)/(C\rho)}$ where $C=\int_{\Re^{2}}g(\left\Vert \boldsymbol{x}\right\Vert )d\boldsymbol{x}$
and $b$ is a constant. Denote the above network by\textmd{ }$\mathcal{G}\left(\mathcal{X}_{\rho},g_{r_{\rho}},A\right)$.
We show that as $\rho\rightarrow\infty$, a) the distribution of the
number of isolated nodes in $\mathcal{G}\left(\mathcal{X}_{\rho},g_{r_{\rho}},A\right)$
converges to a Poisson distribution with mean $e^{-b}$; b) asymptotically
almost surely (\emph{a.a.s.}) there is no component in $\mathcal{G}\left(\mathcal{X}_{\rho},g_{r_{\rho}},A\right)$
of fixed and finite order $k>1$; c) \emph{a.a.s.} the number of components
with an unbounded order is one. Therefore as $\rho\rightarrow\infty$,
the network \emph{a.a.s.} contains a unique unbounded component and
isolated nodes only; a sufficient and necessary condition for $\mathcal{G}\left(\mathcal{X}_{\rho},g_{r_{\rho}},A\right)$
to be \emph{a.a.s.} connected is that there is no isolated node in
the network, which occurs when $b\rightarrow\infty$ as $\rho\rightarrow\infty$.{\normalsize{}
}These results expand recent results obtained for connectivity of
random geometric graphs from the unit disk model and the fewer results
from the log-normal model to the more generic and more practical random
connection model.\textmd{\normalsize{} }{\normalsize \par}\end{abstract}
\begin{IEEEkeywords}
Connectivity, random geometric graph, random connection model
\end{IEEEkeywords}

\section{Introduction\label{sec:Introduction}}

Connectivity is one of the most fundamental properties of wireless
multi-hop networks \cite{Gupta98Critical,Penrose03Random,Haenggi09Stochastic}.
A network is said to be \emph{connected} if there is a path between
any pair of nodes. 

Extensive research has been done on connectivity problems using the
well-known random geometric graph and the unit disk connection model,
which is usually obtained by randomly and uniformly distributing $n$
vertices in a given area and connecting any two vertices iff (if and
only if) their Euclidean distance is smaller than or equal to a given
threshold $r(n)$ \cite{Penrose99On,Penrose03Random}. Significant
outcomes have been obtained \cite{Gupta98Critical,Xue04The,Penrose03Random}.
Particularly, Penrose \cite{Penrose97The,Penrose99A} and Gupta and
Kumar \cite{Gupta98Critical} proved using different techniques that
if the transmission range is set to $r\left(n\right)=\sqrt{(\log n+c\left(n\right))/(\pi n)}$,
a random network formed by uniformly placing $n$ nodes on a unit-area
disk in $\Re^{2}$ is asymptotically almost surely (\emph{a.a.s.})
connected as $n\rightarrow\infty$ iff $c\left(n\right)\rightarrow\infty$.
{[}An event $\xi$ is said to occur \emph{almost surely} if its probability
equals to one; an event $\xi_{n}$ depending on $n$ is said to occur
\emph{a.a.s.} if its probability tends to one as $n\rightarrow\infty${]}.
Specifically, Penrose's result is based on the fact that in the above
random network as $n\rightarrow\infty$ the longest edge of the minimum
spanning tree converges in probability to the minimum transmission
range required for the above network to have no isolated nodes \cite{Penrose97The,Penrose99A,Penrose03Random}.
Gupta and Kumar's result is based on a key finding in continuum percolation
theory \cite[Chapter 6]{Meester96Continuum}: consider an \emph{infinite}
network with nodes distributed on $\Re^{2}$ following a Poisson distribution
with density $\rho$; and suppose that a pair of nodes separated by
an Euclidean distance $x$ are directly connected with probability
$g\left(x\right)$, independent of the event that another distinct
pair of nodes are directly connected. Here, $g:\Re^{+}\rightarrow\left[0,1\right]$
satisfies the conditions of rotational invariance, non-increasing
monotonicity and integral boundedness \cite[pp. 151-152]{Meester96Continuum}.
Denote the above network by $\mathcal{G}\left(\mathcal{X}_{\rho},g,\Re^{2}\right)$.
As $\rho\rightarrow\infty$, \emph{a.a.s.} $\mathcal{G}\left(\mathcal{X}_{\rho},g,\Re^{2}\right)$
has only a unique infinite component and isolated nodes. The work
of Gupta and Kumar is however incomplete to the extent that the above
result obtained in continuum percolation theory for an infinite network
cannot, counter to intuition,  be directly applied to a finite (or
asymptotically infinite) network on a finite (or asymptotically infinite)
area in $\Re^{2}$ \cite{Mao11Towards}. 

In addition to the above work based on the unit disk connection model,
there is also limited work \cite{Bettstetter05Connectivity,Miorandi08The}
dealing with the necessary condition for a random network to be connected
under the log-normal shadowing connection model. Under the log-normal
shadowing connection model, two nodes are directly connected if the
received power at one node from the other node, whose attenuation
follows the log-normal model, is greater than a given threshold. The
results in \cite{Bettstetter05Connectivity,Miorandi08The} however
rely on the\emph{ }assumption that the node isolation events are independent.
This assumption has only been justified using simulations. 

Some work also exists on the analysis of the asymptotic distribution
of the number of isolated nodes \cite{Yi06Asymptotic,Franceschetti06Critical,Franceschetti07Random,Penrose03Random}
under the assumption of a unit disk model. In \cite{Yi06Asymptotic},
Yi et al. considered a total of $n$ nodes distributed independently
and uniformly on a unit-area disk and each node may be active independently
with some probability $p$. A node is considered to be isolated if
it is not directly connected to any of the active nodes. Using some
complicated geometric analysis, they showed that if all nodes have
a maximum transmission range $r(n)=\sqrt{\left(\log n+\xi\right)/(\pi pn)}$
for some constant $\xi$, the total number of isolated nodes is asymptotically
Poissonly distributed with mean $e^{-\xi}$. In \cite{Franceschetti06Critical,Franceschetti07Random},
Franceschetti et al. derived essentially the same result using the
Chen-Stein technique. A similar result can also be found in the earlier
work of Penrose \cite{Penrose03Random} in a continuum percolation
setting. 

In this paper, we consider a network where all nodes are distributed
on a unit square $A\triangleq[-\frac{1}{2},\frac{1}{2}]^{2}$ following
a Poisson distribution with known density $\rho$ and a pair of nodes
are directly connected following a \emph{generic} \emph{random connection
model} $g_{r_{\rho}}$, to be rigorously defined in Section \ref{sec:Network-Model-and-problem-setting}.
Denote the above network by $\mathcal{G}\left(\mathcal{X}_{\rho},g_{r_{\rho}},A\right)$,
where $\mathcal{X}_{\rho}$ denotes the set of nodes in the network.
We give the sufficient and necessary condition for $\mathcal{G}\left(\mathcal{X}_{\rho},g_{r_{\rho}},A\right)$
to be \emph{a.a.s.} connected as $\rho\rightarrow\infty$. The results
in this paper expand the above results on network connectivity to
a more generic random connection model, with the unit disk model and
the log-normal model being two special cases, thus providing an important
link that allows the expansion of other associated results on connectivity
to the random connection model. 

The main contributions of this paper are:
\begin{enumerate}
\item Using the Chen-Stein technique \cite{Arratia90Poisson,Barbour03Poisson},
we show that the distribution of the number of isolated nodes in $\mathcal{G}\left(\mathcal{X}_{\rho},g_{r_{\rho}},A\right)$
asymptotically converges to a Poisson distribution as $\rho\rightarrow\infty$.
This result readily leads to a necessary condition for $\mathcal{G}\left(\mathcal{X}_{\rho},g_{r_{\rho}},A\right)$
to be \emph{a.a.s.} connected as $\rho\rightarrow\infty$;
\item We show that as $\rho\rightarrow\infty$, the number of components
in $\mathcal{G}\left(\mathcal{X}_{\rho},g_{r_{\rho}},A\right)$ of
unbounded order converges to one. This result, together with the result
in \cite{Mao11Towards} that the number of components of finite order
$k>1$ in $\mathcal{G}\left(\mathcal{X}_{\rho},g_{r_{\rho}},A\right)$
asymptotically vanishes as $\rho\rightarrow\infty$, allows us to
conclude that as $\rho\rightarrow\infty$, \emph{a.a.s.} there are
only a unique unbounded component and isolated nodes in $\mathcal{G}\left(\mathcal{X}_{\rho},g_{r_{\rho}},A\right)$. 
\item The above results allow us to establish that the sufficient and necessary
condition for $\mathcal{G}\left(\mathcal{X}_{\rho},g_{r_{\rho}},A\right)$
to be \emph{a.a.s.} connected is that there is no isolated node in
the network. On that basis, we obtain the asymptotic probability that
$\mathcal{G}\left(\mathcal{X}_{\rho},g_{r_{\rho}},A\right)$ forms
a connected network as $\rho\rightarrow\infty$ and the sufficient
and necessary condition for $\mathcal{G}\left(\mathcal{X}_{\rho},g_{r_{\rho}},A\right)$
to be \emph{a.a.s. }connected. 
\end{enumerate}
The rest of this paper is organized as follows: Section \ref{sec:Network-Model-and-problem-setting}
introduces the network model and problem setting; Section \ref{sec:connectivity necessary condition}
establishes a necessary condition for $\mathcal{G}\left(\mathcal{X}_{\rho},g_{r_{\rho}},A\right)$
to be \emph{a.a.s. }connected; Section \ref{sec: connectivity sufficient condition}
first establishes a sufficient condition for $\mathcal{G}\left(\mathcal{X}_{\rho},g_{r_{\rho}},A\right)$
to be \emph{a.a.s. }connected and on that basis, together with the
results in Section \ref{sec:connectivity necessary condition}, then
establishes the sufficient and necessary condition for $\mathcal{G}\left(\mathcal{X}_{\rho},g_{r_{\rho}},A\right)$
to be \emph{a.a.s. }connected; finally Section \ref{sec:Conclusion}
concludes the paper.

\section{Network Model and Problem Setting\label{sec:Network-Model-and-problem-setting}}

We consider a network where all nodes are distributed on a unit square
$A\triangleq[-\frac{1}{2},\frac{1}{2}]^{2}$ following a Poisson distribution
with known density $\rho$ and a pair of nodes are directly connected
following a \emph{random connection model}, viz. a pair of nodes separated
by an Euclidean distance $x$ are directly connected with probability
$g_{r_{\rho}}(x)\triangleq g(x/r_{\rho})$, where $g:\left[0,\infty\right)\rightarrow\left[0,1\right]$,
independent of the event that another pair of nodes are directly connected.
Here 
\begin{equation}
r_{\rho}=\sqrt{(\log\rho+b)/(C\rho)}\label{eq:definition of r_rho}
\end{equation}
and $b$ is a constant. The reason for choosing this particular form
of $r_{\rho}$ is that the analysis becomes nontrivial when $b$ is
a constant. Other forms of $r_{\rho}$ can be accommodated by dropping
the assumption that $b$ is constant, i.e. $b$ becomes a function
of $\rho$, and allowing $b\rightarrow\infty$ or $b\rightarrow-\infty$
as $\rho\rightarrow\infty$. The results are rapidly attainable, and
we discuss these situations separately in Sections \ref{sec:connectivity necessary condition}
and \ref{sec: connectivity sufficient condition}. 

The function $g$ is usually required to satisfy the following properties
of monotonicity, integral boundedness and rotational invariance \cite[Chapter 6]{Franceschetti07Random,Meester96Continuum}%
\footnote{Throughout this paper, we use the non-bold symbol, e.g. $x$, to denote
a scalar and the bold symbol, e.g. $\boldsymbol{x}$, to denote a
vector.%
}:

\begin{eqnarray}
 & g\left(x\right)\leq g\left(y\right) & {\textstyle whenever}\;\; x\geq y\label{eq:conditions on g(x) - non-increasing}\\
 & 0<C\triangleq\int_{\Re^{2}}g(\left\Vert \boldsymbol{x}\right\Vert )d\boldsymbol{x}<\infty\label{eq:conditions on g(x) - integral boundness}
\end{eqnarray}
where $\left\Vert \right\Vert $ represents the Euclidean norm. We
refer readers to \cite[Chapter 6]{Franceschetti07Random,Meester96Continuum}
for detailed discussions on the random connection model.

Equations \eqref{eq:conditions on g(x) - non-increasing} and \eqref{eq:conditions on g(x) - integral boundness}
allow us to conclude that \cite[Equation (3)]{Mao11Towards}%
\footnote{The following notations and definitions are used throughout the paper:
\begin{itemize}
\item $f\left(z\right)=o_{z}(h\left(z\right))$ iff $\lim_{z\rightarrow\infty}\frac{f\left(z\right)}{h\left(z\right)}=0$;
\item $f\left(z\right)=\omega_{z}(h\left(z\right))$ iff $h\left(z\right)=o_{z}\left(f\left(z\right)\right)$;
\item $f\left(z\right)=\Theta_{z}(h\left(z\right))$ iff there exist a sufficiently
large $z_{0}$ and two positive constants $c_{1}$ and $c_{2}$ such
that for any $z>z_{0}$, $c_{1}h\left(z\right)\geq f\left(z\right)\geq c_{2}h\left(z\right)$;
\item $f\left(z\right)\sim_{z}h\left(z\right)$ iff $\lim_{z\rightarrow\infty}\frac{f\left(z\right)}{h\left(z\right)}=1$;\end{itemize}
} 
\begin{equation}
g\left(x\right)=o_{x}(1/x^{2})\label{eq:scaling property of g(x)}
\end{equation}

However, we require $g$ to satisfy the more restrictive requirement
that 
\begin{equation}
g\left(x\right)=o_{x}(1/(x^{2}\log^{2}x))\label{eq:Condition on g(x) requirement 2}
\end{equation}

The condition \eqref{eq:Condition on g(x) requirement 2} is only
slightly more restrictive than \eqref{eq:scaling property of g(x)}
in that for an arbitrarily small positive constant $\varepsilon$,
$1/x^{2+\varepsilon}=o_{x}(1/(x^{2}\log^{2}x))$. The more restrictive
requirement is needed to ensure that the impact of the \emph{truncation
effect} on connectivity is asymptotically vanishingly small as $\rho\rightarrow\infty$
\cite{Mao11Towards}. 

For convenience we also assume that $g$ has infinite support when
necessary. Our results however apply to the situation when $g$ has
bounded support, which forms a special case  and actually makes the
analysis easier.

Denote the above network by $\mathcal{G}\left(\mathcal{X}_{\rho},g_{r_{\rho}},A\right)$.
It is obvious that under a \emph{unit disk model} where $g(x)=1$
for $x\leq1$ and $g(x)=0$ for $x>1$, $r_{\rho}$ corresponds to
the critical transmission range for connectivity \cite{Gupta98Critical}.
Thus the above model incorporates the unit disk model as a special
case. A similar conclusion can also be drawn for the log-normal connection
model.

\section{Necessary Condition for \emph{a.a.s.} Connected Network \label{sec:connectivity necessary condition}}

In this section, as an intermediate step to obtaining the main result,
we first and temporarily consider a network with the same node distribution
and connection model as $\mathcal{G}\left(\mathcal{X}_{\rho},g_{r_{\rho}},A\right)$
however with nodes deployed on a unit torus $A^{T}\triangleq[-\frac{1}{2},\frac{1}{2}]^{2}$.
Denote the network on the torus by $\mathcal{G}^{T}\left(\mathcal{X}_{\rho},g_{r_{\rho}},A\right)$.
We show that as $\rho\rightarrow\infty$, the distribution of the
number of isolated nodes in $\mathcal{G}^{T}\left(\mathcal{X}_{\rho},g_{r_{\rho}},A\right)$,
denoted by $W^{T}$, asymptotically converges to a Poisson distribution
with mean $e^{-b}$. We then extend the above result to $\mathcal{G}\left(\mathcal{X}_{\rho},g_{r_{\rho}},A\right)$.
On that basis, we obtain a necessary condition for $\mathcal{G}\left(\mathcal{X}_{\rho},g_{r_{\rho}},A\right)$
to be \emph{a.a.s.} connected as $\rho\rightarrow\infty$.

\subsection{Distribution of the number of isolated nodes on a torus}

In this subsection, we analyze the distribution of the number of isolated
nodes in $\mathcal{G}^{T}\left(\mathcal{X}_{\rho},g_{r_{\rho}},A\right)$. 

The use of a toroidal rather than planar region as a tool in analyzing
network properties is well known \cite{Penrose03Random}. The unit
torus $A^{T}=[-\frac{1}{2},\frac{1}{2}]^{2}$ that is commonly used
in random geometric graph theory is essentially the same as a unit
square $A=[-\frac{1}{2},\frac{1}{2}]^{2}$ except that the distance
between two points on a torus is defined by their\emph{ toroidal distance},
instead of Euclidean distance. Thus a pair of nodes in $\mathcal{G}^{T}\left(\mathcal{X}_{\rho},g_{r_{\rho}},A\right)$,
located at $\boldsymbol{x}_{1}$ and $\boldsymbol{x}_{2}$ respectively,
are directly connected with probability $g_{r_{\rho}}(\left\Vert \boldsymbol{x}_{1}-\boldsymbol{x}_{2}\right\Vert ^{T})$
where $\left\Vert \boldsymbol{x}_{1}-\boldsymbol{x}_{2}\right\Vert ^{T}$
denotes the \emph{toroidal distance} between the two nodes. For a
unit torus $A^{T}=[-\frac{1}{2},\frac{1}{2}]^{2}$, the toroidal distance
is given by \cite[p. 13]{Penrose03Random}:
\begin{equation}
\left\Vert \boldsymbol{x}_{1}-\boldsymbol{x}_{2}\right\Vert ^{T}\triangleq\min\{\left\Vert \boldsymbol{x}_{1}+\boldsymbol{z}-\boldsymbol{x}_{2}\right\Vert :\boldsymbol{z}\in\mathbb{Z}^{2}\}\label{eq:definition of toroidal distance in a unit torus}
\end{equation}
In this section, whenever the difference between a torus and a square
affects the parameter being discussed, we use superscript $^{T}$
to mark the parameter in a torus while the unmarked parameter is associated
with a square. 

We note the following relation between toroidal distance and Euclidean
distance on a square area centered at the origin:
\begin{eqnarray}
\left\Vert \boldsymbol{x}_{1}-\boldsymbol{x}_{2}\right\Vert ^{T}\leq\left\Vert \boldsymbol{x}_{1}-\boldsymbol{x}_{2}\right\Vert \;\;\textrm{and}\;\;\left\Vert \boldsymbol{x}\right\Vert ^{T}=\left\Vert \boldsymbol{x}\right\Vert \label{eq:property of toroidal distance 1}
\end{eqnarray}
which will be used in the later analysis. 

The main result of this subsection is given in Theorem \ref{thm: Distribution of isolated nodes torus}.
\begin{thm}
\label{thm: Distribution of isolated nodes torus}The distribution
of the number of isolated nodes in \textup{$\mathcal{G}^{T}\left(\mathcal{X}_{\rho},g_{r_{\rho}},A\right)$}
converges to a Poisson distribution with mean $e^{-b}$ as $\rho\rightarrow\infty$.\end{thm}
\begin{IEEEproof}
See Appendix I.
\end{IEEEproof}

\subsection{Distribution of the number of isolated nodes on a square}

We now consider the asymptotic distribution of the number of isolated
nodes in $\mathcal{G}\left(\mathcal{X}_{\rho},g_{r_{\rho}},A\right)$. 

Let $W$ be the number of isolated nodes in \emph{$\mathcal{G}\left(\mathcal{X}_{\rho},g_{r_{\rho}},A\right)$
}and $W^{E}$ be the number of isolated nodes in $\mathcal{G}\left(\mathcal{X}_{\rho},g_{r_{\rho}},A\right)$
due to the boundary effect\emph{. }Using the coupling technique, it
can be readily shown that $W=W^{E}+W^{T}$ \cite{Mao11Towards}. Using
the above equation, Theorem \ref{thm: Distribution of isolated nodes torus},
Lemma 2 in \cite{Mao11Towards}%
\footnote{Let $\mathcal{G}\left(\mathcal{X}_{\lambda},g,A_{\frac{1}{r_{\rho}}}\right)$
be a network with nodes Poissonly distributed on a square $A_{\frac{1}{r_{\rho}}}=[-\frac{1}{2r_{\rho}},\frac{1}{2r_{\rho}}]^{2}$
with density $\lambda=(\log\rho+b)/C$ and a pair of nodes separated
by an Euclidean distance $x$ are directly connected with probability
$g(x)$, independent of other connections. Results in \cite{Mao11Towards}
are derived for $\mathcal{G}\left(\mathcal{X}_{\lambda},g,A_{\frac{1}{r_{\rho}}}\right)$.
By proper scaling, it is straightforward to extend the results for
$\mathcal{G}\left(\mathcal{X}_{\lambda},g,A_{\frac{1}{r_{\rho}}}\right)$
to $\mathcal{G}\left(\mathcal{X}_{\rho},g_{r_{\rho}},A\right)$. Therefore
we ignore the difference.%
}, which showed that $\lim_{\rho\rightarrow\infty}\Pr(W^{E}=0)=1$,
and Slutsky's theorem \cite{Grimmett01Probability}, the following
result on the asymptotic distribution of $W$ can be readily obtained.
\begin{thm}
\label{thm:Number of isolated nodes square}The distribution of the
number of isolated nodes in \textup{$\mathcal{G}\left(\mathcal{X}_{\rho},g_{r_{\rho}},A\right)$}
converges to a Poisson distribution with mean $e^{-b}$ as $\rho\rightarrow\infty$.
\end{thm}
Corollary \ref{cor:Prob no isolated node} follows immediately from
Theorem \ref{thm:Number of isolated nodes square}.
\begin{cor}
\label{cor:Prob no isolated node}As $\rho\rightarrow\infty$, the
probability that there is no isolated node in \textup{$\mathcal{G}\left(\mathcal{X}_{\rho},g_{r_{\rho}},A\right)$}
converges to $e^{-e^{-b}}$.
\end{cor}
Now we relax requirement that $b$ is a constant to obtain a necessary
condition for $\mathcal{G}\left(\mathcal{X}_{\rho},g_{r_{\rho}},A\right)$
to be \emph{a.a.s.} connected. Specifically, consider the situation
when $b\rightarrow-\infty$ or $b\rightarrow\infty$ as $\rho\rightarrow\infty$.
Note that the property that the network $\mathcal{G}\left(\mathcal{X}_{\rho},g_{r_{\rho}},A\right)$
has no isolated node is an \emph{increasing} property (For an arbitrary
network, a particular property is termed \emph{increasing} if the
property is preserved when more connections (edges) are added into
the network.). Using a coupling technique similar to that used in
\cite[Chapter 2]{Franceschetti07Random} and with a few simple steps
(omitted), the following theorem and corollary can be obtained, which
form a major contribution of this paper:
\begin{thm}
\label{thm:isolated nodes b=00003D0 or infinity}In \textup{$\mathcal{G}\left(\mathcal{X}_{\rho},g_{r_{\rho}},A\right)$},
if $b\rightarrow\infty$ as $\rho\rightarrow\infty$, a.a.s. there
is no isolated node in the network; if $b\rightarrow-\infty$ as $\rho\rightarrow\infty$,
a.a.s.\emph{ }the network has at least one isolated node.\end{thm}
\begin{cor}
\label{cor:necessary condition for asymptotically connected network}$b\rightarrow\infty$
is a necessary condition for \textup{$\mathcal{G}\left(\mathcal{X}_{\rho},g_{r_{\rho}},A\right)$}
to be a.a.s. connected as $\rho\rightarrow\infty$.
\end{cor}

\section{Sufficient Condition for \emph{a.a.s.} Connected Network\label{sec: connectivity sufficient condition}}

In this section, we continue to investigate the sufficient condition
for $\mathcal{G}\left(\mathcal{X}_{\rho},g_{r_{\rho}},A\right)$ to
be \emph{a.a.s.} connected. In \cite{Mao11Towards} we showed that
vanishing of components of finite order $k>1$ in $\mathcal{G}\left(\mathcal{X}_{\rho},g,\Re^{2}\right)$
as $\rho\rightarrow\infty$ (as shown in \cite[Theorems 6.3]{Meester96Continuum})
does not \emph{necessarily} carry the conclusion that components of
finite order $k>1$ in $\mathcal{G}\left(\mathcal{X}_{\rho},g_{r_{\rho}},A\right)$
also vanish as $\rho\rightarrow\infty$, contrary perhaps to intuition.
Then, we presented a result for the vanishing of components of finite
order $k>1$ in $\mathcal{G}\left(\mathcal{X}_{\rho},g_{r_{\rho}},A\right)$
as $\rho\rightarrow\infty$ to fill this theoretical gap \cite[Theorem 4]{Mao11Towards}.
On the basis of the above results, we shall further demonstrate in
this section that \emph{a.a.s.} the number of unbounded components
in $\mathcal{G}\left(\mathcal{X}_{\rho},g_{r_{\rho}},A\right)$ is
one as $\rho\rightarrow\infty$. A sufficient condition for $\mathcal{G}\left(\mathcal{X}_{\rho},g_{r_{\rho}},A\right)$
to be \emph{a.a.s.} connected readily follows.

In \cite[Theorem 6.3]{Meester96Continuum}, it was shown that there
can be at most one unbounded component in $\mathcal{G}\left(\mathcal{X}_{\rho},g,\Re^{2}\right)$.
However due to the truncation effect \cite{Mao11Towards}, it appears
difficult to establish such a conclusion using \cite[Theorem 6.3]{Meester96Continuum}.
Indeed differently from $\mathcal{G}\left(\mathcal{X}_{\rho},g,\Re^{2}\right)$
in which an unbounded component may exist for a finite $\rho$, it
can be easily shown that for any finite $\rho$, $\Pr\left(\left|\mathcal{X}_{\rho}\right|<\infty\right)=1$,
i.e. the total number of nodes in $\mathcal{G}\left(\mathcal{X}_{\rho},g_{r_{\rho}},A\right)$
is almost surely finite. It then follows that for any finite $\rho$
almost surely there is no unbounded component in $\mathcal{G}\left(\mathcal{X}_{\rho},g_{r_{\rho}},A\right)$
. 

In this paper, we solve the above conceptual difficulty involving
use of the term ``unbounded component'' by considering the number
of components in $\mathcal{G}\left(\mathcal{X}_{\rho},g_{r_{\rho}},A\right)$
of order greater than $M$, denoted by $\xi_{>M}$, where $M$ is
an arbitrarily large positive integer. We then show that $\lim_{M\rightarrow\infty}\lim_{\rho\rightarrow\infty}\Pr(\xi_{>M}=1)=1$.
The analytical result is summarized in the following theorem, which
forms a further major contribution of this paper:
\begin{thm}
\label{thm:component of unbounded order}As $\rho\rightarrow\infty$,
\emph{a.a.s.} the number of unbounded components in $\mathcal{G}\left(\mathcal{X}_{\rho},g_{r_{\rho}},A\right)$
is one.\end{thm}
\begin{IEEEproof}
See Appendix II\end{IEEEproof}
\begin{rem}
Proof of the type of results in Theorem \ref{thm:component of unbounded order}
usually requires some complicated geometric analysis. Particularly
the proof of Lemma \ref{lem:convergence of X to Poisson} in Appendix
II, which forms a foundation of the proof of Theorem \ref{thm:component of unbounded order},
needs sophisticated geometric analysis. In this paper, we omitted
the proof of Lemma \ref{lem:convergence of X to Poisson} because
the proof is exactly the same as the proof of Theorem \ref{thm:Number of isolated nodes square},
which in turn relies on some results established in \cite{Mao11Towards}.
We refer interested readers to the proof of Theorem 1 in \cite{Mao11Towards}
for techniques on handling geometric obstacles involved in analyzing
the boundary effect and to the proof of Theorem 4 in \cite{Mao11Towards}
for techniques on handling geometric obstacles involved in analyzing
the number of components in $\mathcal{G}\left(\mathcal{X}_{\rho},g_{r_{\rho}},A\right)$.
\end{rem}
An implication of Theorem \ref{thm:component of unbounded order}
is that for an arbitrarily small positive constant $\varepsilon$,
there exists large positive constants $M_{0}$ and $\rho_{0}$ such
that for all $M>M_{0}$ and $\rho>\rho_{0}$, $\Pr(\xi_{>M}=1)>1-\varepsilon$.
From \eqref{eq:convergence of number of components of order greater than M}
in Appendix II, it can further be concluded that for a particular
positive integer $M$ and an arbitrarily small positive constant $\varepsilon$,
there exists $\rho_{0}$ such that for all $\rho>\rho_{0}$, 
\begin{equation}
\Pr(\xi_{>M}=1)>1-\frac{e^{-\left(M+1\right)b}}{\left(M+1\right)!}-\varepsilon\label{eq:result on vanishing of component of finite order}
\end{equation}

The following corollary can be obtained from \cite[Theorem 4]{Mao11Towards}
and Theorem \ref{thm:component of unbounded order}:
\begin{cor}
\label{cor:sufficient and necessary condition for connectivity}As
$\rho\rightarrow\infty$, a.a.s.\emph{ }$\mathcal{G}\left(\mathcal{X}_{\rho},g_{r_{\rho}},A\right)$
forms a connected network iff there is no isolated node in it.\end{cor}
\begin{IEEEproof}
Let $\xi$ be the total number of components in $\mathcal{G}\left(\mathcal{X}_{\rho},g_{r_{\rho}},A\right)$.
It is clear that $\xi=\xi_{1}+\sum_{k=2}^{M}\xi_{k}+\xi_{>M}$, where
$\xi_{k}$ is the number of components of order $k$. Noting that
$\xi=1$ iff $\mathcal{G}\left(\mathcal{X}_{\rho},g_{r_{\rho}},A\right)$
forms a connected network, it suffices to show that $\lim_{\rho\rightarrow\infty}\Pr\left(\xi=1|\xi_{1}=0\right)=1$.$ $
We observe that

\begin{align}
 & \Pr\left(\xi=1,\xi_{1}=0\right)\nonumber \\
\geq & \Pr(\xi_{1}=0,\sum_{k=2}^{M}\xi_{k}=0,\xi_{>M}=1)\nonumber \\
= & \Pr(\xi_{1}=0)-(\Pr(\overline{\sum_{k=2}^{M}\xi_{k}=0})+\Pr(\overline{\xi_{>M}=1}))\label{eq:final result on the joint probability}
\end{align}
where in \eqref{eq:final result on the joint probability} $\overline{\xi_{>M}=1}$
represents the complement of the event $\xi_{>M}=1$ and \eqref{eq:final result on the joint probability}
results as a consequence of the union bound. Further note that \eqref{eq:final result on the joint probability}
is valid for \emph{any} value of $M$ and that $\Pr\left(\xi_{1}=0\right)$
converges to a \emph{non-zero} constant $e^{-e^{-b}}$ as $\rho\rightarrow\infty$
(Theorem \ref{thm:Number of isolated nodes square}). Using the above
results, \cite[Theorem 4]{Mao11Towards} which showed that $\lim_{\rho\rightarrow\infty}\Pr(\sum_{k=2}^{M}\xi_{k}=0)=1$,
and \eqref{eq:result on vanishing of component of finite order},
and following a few simple steps (omitted), it can be shown that for
an arbitrarily small positive constant $\varepsilon$, by choosing
$M$ to be sufficiently large, there exists $\rho_{0}$ such that
for all $\rho>\rho_{0}$, $\Pr\left(\xi=1|\xi_{1}=0\right)>1-\varepsilon$.
\end{IEEEproof}
As an easy consequence of Theorem \ref{thm:Number of isolated nodes square}
and Corollary \ref{cor:sufficient and necessary condition for connectivity},
the following theorem can be established:
\begin{thm}
As $\rho\rightarrow\infty$, the probability that $\mathcal{G}\left(\mathcal{X}_{\rho},g_{r_{\rho}},A\right)$
forms a connected network converges to $e^{-e^{-b}}$.
\end{thm}
Using the above theorem and a similar analysis as that leading to
Theorem \ref{thm:isolated nodes b=00003D0 or infinity} and Corollary
\ref{cor:necessary condition for asymptotically connected network},
the following theorem on the sufficient and necessary condition for
$\mathcal{G}\left(\mathcal{X}_{\rho},g_{r_{\rho}},A\right)$ to be
\emph{a.a.s.} connected can be obtained:
\begin{thm}
\label{thm: sufficient and necessary condition} As $\rho\rightarrow\infty$,
$\mathcal{G}\left(\mathcal{X}_{\rho},g_{r_{\rho}},A\right)$ is \emph{a.a.s.}
connected iff $b\rightarrow\infty$; $\mathcal{G}\left(\mathcal{X}_{\rho},g_{r_{\rho}},A\right)$
is \emph{a.a.s. }disconnected iff $b\rightarrow-\infty$.
\end{thm}

\section{Conclusion and Further Work\label{sec:Conclusion}}

Following the seminal work of Penrose \cite{Penrose99On,Penrose03Random}
and Gupta and Kumar \cite{Gupta98Critical} on the asymptotic connectivity
of large-scale random networks with Poisson node distribution and
under the unit disk model, there is general expectation that there
is a range of connection functions for which the above results \cite{Penrose99On,Penrose03Random,Gupta98Critical}
obtained assuming the unit disk model can carry over. However, for
quite a long time, both the asymptotic laws that the network should
follow and the conditions on the connection function required for
the network to be \emph{a.a.s.} connected under a more generic setting
have been unknown. In this paper, we filled in the gaps by providing
the sufficient and necessary condition for a network with nodes Poissonly
distributed on a unit square and following a generic random connection
model to be \emph{a.a.s.} connected as $\rho\rightarrow\infty$. The
conditions on the connection function required in order for the above
network to be \emph{a.a.s.} connected were also provided. Therefore,
the results in the paper constitute a significant advance of the earlier
work by Penrose \cite{Penrose99On,Penrose03Random} and Gupta and
Kumar \cite{Gupta98Critical} from the unit disk model to the more
generic random connection model and bring models addressed by theoretical
research closer to reality. 

However, there remain significant challenges ahead. The results in
this paper rely on three main assumptions: a) the connection function
$g$ is isotropic, b) the random events underpinning generation of
a connection are independent, c) nodes are Poissonly distributed.
We conjecture that assumption a) is not a critical assumption, i.e.
under some mild conditions, e.g. nodes are independently and randomly
oriented, assumption a) can be removed while our results are still
valid. It is part of our future work plan to validate the conjecture.
Our results however critically rely on assumption b), which is not
necessarily valid in some real networks due to channel correlation
and interference, where the latter effect makes the connection between
a pair of nodes dependent on the locations and activities of other
nearby nodes. In \cite{Yang12Connectivity} we have done some preliminary
work on network connectivity considering the impact of interference.
The work essentially uses a de-coupling approach to solve the challenges
of connection correlation caused by interference and suggests that
when some realistic constraints are considered, i.e. carrier-sensing,
the connectivity results will be very close to those obtained under
a unit disk model. This conclusion is in contrast with that \cite{Dousse05Impact}
obtained under an ALOHA multiple-access protocol. A more thorough
investigation is yet to be done. The major obstacle in dealing with
the impact of channel correlation is that there is no widely accepted
model in the wireless communication community capturing the impact
of channel correlation on connections. Finally, it is a logical move
after our work to consider connectivity of networks with nodes following
a generic distribution other than Poisson. It is part of our future
work plan to tackle the problem.

\section*{Appendix I: Proof of Theorem \ref{thm: Distribution of isolated nodes torus}}

Our proof relies on the use of the Chen-Stein bound \cite{Arratia90Poisson,Barbour03Poisson}.
We first establish some preliminary results that allow us to use the
Chen-Stein bound for the analysis of number of isolated nodes in $\mathcal{G}^{T}\left(\mathcal{X}_{\rho},g_{r_{\rho}},A\right)$. 

Divide the unit torus into $m^{2}$ non-overlapping squares each with
size $\frac{1}{m^{2}}$. Denote the $i_{m}^{th}$ square by $A_{i_{m}}$.
Define two sets of indicator random variables $J_{i_{m}}^{T}$ and
$I_{i_{m}}^{T}$ with $i_{m}\in\Gamma_{m}\triangleq\{1,\ldots m^{2}\}$,
where $J_{i_{m}}^{T}=1$ iff there exists exactly one node in $A_{i_{m}}$,
otherwise $J_{i_{m}}^{T}=0$; $I_{i_{m}}^{T}=1$ iff there is exactly
one node in $A_{i_{m}}$ \emph{and} that node is isolated, $I_{i_{m}}^{T}=0$
otherwise. Obviously $J_{i_{m}}^{T}$ is independent of $J_{j_{m}}^{T},j_{m}\in\Gamma_{m}\backslash\left\{ i_{m}\right\} $.
Denote the center of $A_{i_{m}}^{T}$ by $\boldsymbol{x}_{i_{m}}$
and without loss of generality we assume that when $J_{i_{m}}^{T}=1$,
the associated node in $A_{i_{m}}$ is at $\boldsymbol{x}_{i_{m}}$%
\footnote{In this paper we are mainly concerned with the case that $m\rightarrow\infty$,
i.e. the size of the square is vanishingly small. Therefore the actual
position of the node in the square is not important.%
}. Observe that for any fixed $m$, the values of $\Pr\left(I_{i_{m}}^{T}=1\right)$
and $\Pr\left(J_{i_{m}}^{T}=1\right)$ do not depend on the particular
index $i_{m}$ on a torus. However both the set of indices $\Gamma_{m}$
and a particular index $i_{m}$ depend on $m$. As $m$ changes, the
square associated with $I_{i_{m}}^{T}$ and $J_{i_{m}}^{T}$ also
changes. 
\begin{rem}
In this paper, we are only interested in the limiting values of various
parameters associated with a sub-square as $m\rightarrow\infty$.
Also because of the consideration of a torus, the value of a particular
index $i_{m}$ does not affect the discussion of the associated parameters,
i.e. these parameters $I_{i_{m}}^{T}$ and $J_{i_{m}}^{T}$ do not
depend on $i_{m}$. Therefore in the following, we omit some straightforward
discussions on the convergence of various parameters, e.g. $i_{m}$,
$\boldsymbol{x}_{i_{m}}$, $I_{i_{m}}^{T}$ and $J_{i_{m}}^{T}$,
as $m\rightarrow\infty$.
\end{rem}
Without causing ambiguity, we drop the explicit dependence on $m$
in our notations for convenience. As an easy consequence of the Poisson
node distribution, $\Pr(J_{i}^{T}=1)\sim_{m}\rho/m^{2}$. Using \cite[Proposition 1.3]{Meester96Continuum},
$\Pr(I_{i}^{T}=1)=\Pr(I_{i}^{T}=1|J_{i}^{T}=1)\Pr(J_{i}^{T}=1)$ and
the property of a torus (see also \cite[Lemma 1]{Mao11Towards}),
it can be shown that 
\begin{eqnarray}
\Pr(I_{i}^{T}=1) & \sim_{m} & \frac{\rho}{m^{2}}e^{-\int_{A}\rho g(\frac{\left\Vert \boldsymbol{x}-\boldsymbol{x}_{i}\right\Vert ^{T}}{r_{\rho}})d\boldsymbol{x}}\nonumber \\
 & = & \frac{\rho}{m^{2}}e^{-\int_{A}\rho g(\frac{\left\Vert \boldsymbol{x}\right\Vert ^{T}}{r_{\rho}})d\boldsymbol{x}}\label{eq:prob isolated node}
\end{eqnarray}

Now consider the event $I_{i}^{T}I_{j}^{T}=1,i\neq j$, conditioned
on the event that $J_{i}^{T}J_{j}^{T}=1$, meaning that both nodes
having been placed inside $A_{i}$ and $A_{j}$ respectively are isolated.
Following the same steps leading to \eqref{eq:prob isolated node},
it can be shown that

\begin{align}
 & \lim_{m\rightarrow\infty}\Pr(I_{i}^{T}I_{j}^{T}=1|J_{i}^{T}J_{j}^{T}=1)\nonumber \\
= & (1-g(\frac{\left\Vert \boldsymbol{x}_{i}-\boldsymbol{x}_{j}\right\Vert ^{T}}{r_{\rho}}))\exp[-\int_{A}\rho(g(\frac{\left\Vert \boldsymbol{x}-\boldsymbol{x}_{i}\right\Vert ^{T}}{r_{\rho}})\nonumber \\
+ & g(\frac{\left\Vert \boldsymbol{x}-\boldsymbol{x}_{j}\right\Vert ^{T}}{r_{\rho}})-g(\frac{\left\Vert \boldsymbol{x}-\boldsymbol{x}_{i}\right\Vert ^{T}}{r_{\rho}})g(\frac{\left\Vert \boldsymbol{x}-\boldsymbol{x}_{j}\right\Vert ^{T}}{r_{\rho}}))d\boldsymbol{x}]\label{eq:joint distribution of isolated node events}
\end{align}
where the term $(1-g(\frac{\left\Vert \boldsymbol{x}_{i}-\boldsymbol{x}_{j}\right\Vert ^{T}}{r_{\rho}}))$
is due to the requirement that the two nodes located inside $A_{i}$
and $A_{j}$ cannot be directly connected given that they are both
isolated nodes. Observe also that $\Pr(I_{i}^{T}I_{j}^{T}=1)=\Pr(J_{i}^{T}J_{j}^{T}=1)\Pr(I_{j}^{T}I_{j}^{T}=1|J_{i}^{T}J_{j}^{T}=1)$.
Now using the above equation, \eqref{eq:prob isolated node} and \eqref{eq:joint distribution of isolated node events},
it can be established that
\begin{align}
 & \frac{\Pr(I_{i}^{T}I_{j}^{T}=1)}{\Pr(I_{i}^{T}=1)\Pr(I_{j}^{T}=1)}\nonumber \\
\sim_{m} & (1-g(\frac{\left\Vert \boldsymbol{x}_{i}-\boldsymbol{x}_{j}\right\Vert ^{T}}{r_{\rho}}))e^{\int_{A}\rho g(\frac{\left\Vert \boldsymbol{x}-\boldsymbol{x}_{i}\right\Vert ^{T}}{r_{\rho}})g(\frac{\left\Vert \boldsymbol{x}-\boldsymbol{x}_{j}\right\Vert ^{T}}{r_{\rho}})d\boldsymbol{x}}\label{eq:ratio correlation}
\end{align}

Now we are ready to use the Chen-Stein bound to prove Theorem \ref{thm: Distribution of isolated nodes torus}.
Particularly, we will show using the Chen-Stein bound that 
\begin{equation}
W^{T}=\lim_{m\rightarrow\infty}\sum_{i\in\Gamma_{m}}I_{i}^{T}\label{eq: relating number of isolated nodes to indicators}
\end{equation}
asymptotically converges to a Poisson distribution with mean $e^{-b}$
as $\rho\rightarrow\infty$. 

The following theorem gives a formal statement of the Chen-Stein bound:
\begin{thm}
\label{thm:Chen-Stein Bound}\cite[Theorem 1.A]{Barbour03Poisson}
For a set of indicator random variables $I_{i},\; i\in\Gamma$, define
$W\triangleq\sum_{i\in\Gamma}I_{i}$, $p_{i}\triangleq E\left(I_{i}\right)$
and $\eta\triangleq E\left(W\right)$. For any choice of the index
set $\Gamma_{s,i}\subset\Gamma$, $\Gamma_{s,i}\cap\{i\}=\{\textrm{�}\}$,
\begin{eqnarray*}
 &  & d_{TV}(\mathcal{L}\left(W\right),Po\left(\eta\right))\\
 & \leq & \sum_{i\in\Gamma}[(p_{i}^{2}+p_{i}E(\sum_{j\in\Gamma_{s,i}}I_{j}))]\min(1,\frac{1}{\eta})\\
 & + & \sum_{i\in\Gamma}E(I_{i}\sum_{j\in\Gamma_{s,i}}I_{j})\min(1,\frac{1}{\eta})\\
 & + & \sum_{i\in\Gamma}E|E\{I_{i}|(I_{j},j\in\Gamma_{w,i})\}-p_{i}|\min(1,\frac{1}{\eta})
\end{eqnarray*}
where $\mathcal{L}\left(W\right)$ denotes the distribution of $W$,
$Po\left(\eta\right)$ denotes a Poisson distribution with mean $\eta$,
$\Gamma_{w,i}=\Gamma\backslash\left\{ \Gamma_{s,i}\cup\{i\}\right\} $
and $d_{TV}$ denotes the total variation distance. The total variation
distance between two probability distributions $\alpha$ and $\beta$
on $\mathbb{Z}^{+}$ is given by $d_{TV}\left(\alpha,\beta\right)\triangleq\sup\left\{ \left|\alpha\left(A\right)-\beta\left(A\right)\right|:A\subset\mathbb{Z}^{+}\right\} $.
\end{thm}
For convenience, we separate the bound in Theorem \ref{thm:Chen-Stein Bound}
into three terms $b_{1}\min(1,\frac{1}{\eta})$, $b_{2}\min(1,\frac{1}{\eta})$
and $b_{3}\min(1,\frac{1}{\eta})$ where

\begin{align}
b_{1}\triangleq & \sum_{i\in\Gamma}[(p_{i}^{2}+p_{i}E(\sum_{j\in\Gamma_{s,i}}I_{j}))]\label{eq:definition of b1}\\
b_{2}\triangleq & \sum_{i\in\Gamma}E(I_{i}\sum_{j\in\Gamma_{s,i}}I_{j})\label{eq:definition of b2}\\
b_{3}\triangleq & \sum_{i\in\Gamma}E|E\{I_{i}|(I_{j},j\in\Gamma_{w,i})\}-p_{i}|\label{eq:definition of b3}
\end{align}

The set of indices $\Gamma_{s,i}$ is often chosen to contain all
those $j$, other than $i$, for which $I_{j}$ is ``strongly''
dependent on $I_{i}$ and the set $\Gamma_{w,i}$ often contains all
other indices apart from $i$ for which $I_{j}$ is at most ``weakly''
dependent on $I_{i}$ \cite{Arratia90Poisson}. 
\begin{rem}
A main challenge in using the Chen-Stein bound to prove Theorem \ref{thm: Distribution of isolated nodes torus}
is that under the random connection model, the two events $I_{i}$
and $I_{j}$ may be correlated even when $\boldsymbol{x}_{i}$ and
$\boldsymbol{x}_{j}$ are separated by a very large Euclidean distance.
Therefore the dependence structure is global, which significantly
increases the complexity of the analysis. In comparison, in applications
where the dependence structure is local, by a suitable choice of $\Gamma_{s,i}$
the $b_{3}$ term can be easily made to be $0$ and the evaluation
of the $b_{1}$ and $b_{2}$ terms involves the computation of the
first two moments of $W$ only, which can often be achieved relatively
easily. An example is a random geometric network under the unit disk
model. If $\Gamma_{s,i}$ is chosen to be a neighborhood of $i$ containing
indices of all nodes whose distance to node $i$ is less than or equal
to twice the transmission range, the $b_{3}$ term is easily shown
to be $0$. It can then be readily shown that the $b_{1}$ and $b_{2}$
terms approach $0$ as the neighborhood size of a node becomes vanishingly
small compared to the overall network size as $\rho\rightarrow\infty$
\cite{Franceschetti06Critical}. However this is certainly not the
case for the random connection model. 
\end{rem}

\begin{rem}
The key idea involved using the Chen-Stein bound to prove Theorem
\ref{thm: Distribution of isolated nodes torus} is constructing a
neighborhood of a node, i.e. $\Gamma_{s,i}$ in Theorem \ref{thm:Chen-Stein Bound},
such that a) the size of the neighborhood becomes vanishingly small
compared with $A$ as $\rho\rightarrow\infty$. This is required for
the $b_{1}$ and $b_{2}$ terms to approach $0$ as $\rho\rightarrow\infty$;
b) \emph{a.a.s.} the neighborhood contains all nodes that may have
a direct connection with the node. This is required for the $b_{3}$
term to approach $0$ as $\rho\rightarrow\infty$. Such a neighborhood
is defined in the next paragraph.
\end{rem}
Let $D^{T}\left(\boldsymbol{x}_{i},r\right)\triangleq\{\boldsymbol{x}\in A:\left\Vert \boldsymbol{x}-\boldsymbol{x}_{i}\right\Vert ^{T}\leq r\}$
and when $\boldsymbol{x}_{i}$ is not within $r$ of the border of
$A$, $ $$D^{T}\left(\boldsymbol{x}_{i},r\right)$ becomes the same
as $D\left(\boldsymbol{x}_{i},r\right)$ where $D\left(\boldsymbol{x}_{i},r\right)\triangleq\{\boldsymbol{x}\in A:\left\Vert \boldsymbol{x}-\boldsymbol{x}_{i}\right\Vert \leq r\}$.
Further define the neighborhood of an index $i\in\Gamma$ as $\Gamma_{s,i}\triangleq\{j:\boldsymbol{x}_{j}\in D^{T}\left(\boldsymbol{x}_{i},2r_{\rho}^{1-\epsilon}\right)\}\backslash\{i\}$
and define the non-neighborhood of the index $i$ as $\Gamma_{w,i}\triangleq\{j:\boldsymbol{x}_{j}\notin D^{T}(\boldsymbol{x}_{i},2r_{\rho}^{1-\epsilon})\}$
where $\epsilon$ is a small positive constant and $\epsilon\in(0,\frac{1}{2})$.
It can be shown that

\begin{equation}
\left|\Gamma_{s,i}\right|=m^{2}4\pi r_{\rho}^{2-2\epsilon}+o_{m}(m^{2}4\pi r_{\rho}^{2-2\epsilon})\label{eq:tao s, i}
\end{equation}

Note that in Theorem \ref{thm:Chen-Stein Bound}, $p_{i}=E(I_{i}^{T})$
and $E(I_{i}^{T})$ has been given in \eqref{eq:prob isolated node}.
Further, as an easy consequence of \eqref{eq: relating number of isolated nodes to indicators}
and \cite[Lemma 1]{Mao11Towards} which showed that 
\begin{equation}
\lim_{\rho\rightarrow\infty}E(W^{T})=\lim_{\rho\rightarrow\infty}\rho e^{-\int_{A}\rho g(\frac{\left\Vert \boldsymbol{x}\right\Vert ^{T}}{r_{\rho}})d\boldsymbol{x}}=e^{-b}\label{eq:expected number of isolatde nodes torus}
\end{equation}
 $\lim_{\rho\rightarrow\infty}\lim_{m\rightarrow\infty}\eta=e^{-b}$. 

Using \eqref{eq:prob isolated node}, $p_{i}=E(I_{i}^{T})$ and \eqref{eq:expected number of isolatde nodes torus},
it follows that
\begin{eqnarray}
\lim_{m\rightarrow\infty}m^{2}p_{i} & = & \rho e^{-\int_{A}\rho g(\frac{\left\Vert \boldsymbol{x}\right\Vert ^{T}}{r_{\rho}})d\boldsymbol{x}}\label{eq:value of m2p_i finite rho}\\
\lim_{\rho\rightarrow\infty}\lim_{m\rightarrow\infty}m^{2}p_{i} & = & e^{-b}\label{eq:limiting value m2p_i}
\end{eqnarray}

Next we shall evaluate the $b_{1}$, $b_{2}$ and $b_{3}$ terms in
the following three subsections separately and show that all three
terms converge to $0$ as $\rho\rightarrow\infty$.

\subsection{An Evaluation of the $b_{1}$ Term\label{sub:An-Evaluation-of-b1}}

It can be shown that (following the equation, detailed explanations
are given)

\begin{align}
 & \lim_{\rho\rightarrow\infty}\lim_{m\rightarrow\infty}\sum_{i\in\Gamma}(p_{i}^{2}+p_{i}E(\sum_{j\in\Gamma_{s,i}}I_{j}^{T}))\nonumber \\
= & \lim_{\rho\rightarrow\infty}\lim_{m\rightarrow\infty}m^{2}p_{i}E(\sum_{j\in\Gamma_{s,i}\cup\left\{ i\right\} }I_{j}^{T})\nonumber \\
= & \lim_{\rho\rightarrow\infty}\lim_{m\rightarrow\infty}(m^{2}p_{i})^{2}4\pi r_{\rho}^{2-2\epsilon}\label{eq:evaluation of b1 second step}\\
= & \lim_{\rho\rightarrow\infty}4\pi(\rho e^{-\int_{A}\rho g(\frac{\left\Vert \boldsymbol{x}-\boldsymbol{x}_{i}\right\Vert ^{T}}{r_{\rho}})d\boldsymbol{x}})^{2}(\frac{\log\rho+b}{C\rho})^{1-\epsilon}\label{eq:evaluation-b1-finite-rho}\\
= & 4\pi e^{-2b}\lim_{\rho\rightarrow\infty}(\frac{\log\rho+b}{C\rho})^{1-\epsilon}=0\label{eq:evaluation-b1-final}
\end{align}
where \eqref{eq:tao s, i} is used in obtaining \eqref{eq:evaluation of b1 second step};
\eqref{eq:definition of r_rho} and \eqref{eq:value of m2p_i finite rho}
are used in obtaining \eqref{eq:evaluation-b1-finite-rho}; and \eqref{eq:expected number of isolatde nodes torus}
and \eqref{eq:limiting value m2p_i} are used in obtaining \eqref{eq:evaluation-b1-final}.
Therefore $\lim_{\rho\rightarrow\infty}\lim_{m\rightarrow\infty}b_{1}=0$.

\subsection{An Evaluation of the $b_{2}$ Term\label{sub:An-Evaluation-of-b2}}

For the $b_{2}$ term, assume that $\rho$ is sufficiently large such
that $\frac{1}{2r_{\rho}}>>2r_{\rho}^{-\epsilon}$ and let $A_{\frac{1}{r_{\rho}}}=[-\frac{1}{2r_{\rho}},\frac{1}{2r_{\rho}}]^{2}$.
Using \eqref{eq:joint distribution of isolated node events} in the
first step; and first using some translation and scaling operations
and then using \eqref{eq:property of toroidal distance 1} in the
last step, equation \eqref{eq:Value-of-b2-finite-rho} can be obtained.
\begin{figure*}[!tbh]
\begin{align}
 & \lim_{m\rightarrow\infty}\sum_{i\in\Gamma}E(I_{i}^{T}\sum_{j\in\Gamma_{s,i}}I_{j}^{T})\nonumber \\
= & \lim_{m\rightarrow\infty}\frac{\rho^{2}}{m^{2}}\sum_{j\in\Gamma_{s,i}}\{(1-g(\left\Vert \frac{\boldsymbol{x}_{i}-\boldsymbol{x}_{j}}{r_{\rho}}\right\Vert ^{T})\exp[-\int_{A}\rho(g(\left\Vert \frac{\boldsymbol{x}-\boldsymbol{x}_{i}}{r_{\rho}}\right\Vert ^{T})+g(\left\Vert \frac{\boldsymbol{x}-\boldsymbol{x}_{j}}{r_{\rho}}\right\Vert ^{T})-g(\left\Vert \frac{\boldsymbol{x}-\boldsymbol{x}_{i}}{r_{\rho}}\right\Vert ^{T})g(\left\Vert \frac{\boldsymbol{x}-\boldsymbol{x}_{j}}{r_{\rho}}\right\Vert ^{T}))])\}d\boldsymbol{x}\nonumber \\
= & \rho^{2}\int_{D^{T}\left(\boldsymbol{x}_{i},2r_{\rho}^{1-\epsilon}\right)}\{(1-g(\frac{\left\Vert \boldsymbol{x}_{i}-\boldsymbol{y}\right\Vert ^{T}}{r_{\rho}}))\exp[-\int_{A}\rho(g(\left\Vert \frac{\boldsymbol{x}-\boldsymbol{x}_{i}}{r_{\rho}}\right\Vert ^{T})+g(\left\Vert \frac{\boldsymbol{x}-\boldsymbol{y}}{r_{\rho}}\right\Vert ^{T})-g(\left\Vert \frac{\boldsymbol{x}-\boldsymbol{x}_{i}}{r_{\rho}}\right\Vert ^{T})g(\left\Vert \frac{\boldsymbol{x}-\boldsymbol{y}}{r_{\rho}}\right\Vert ^{T}))d\boldsymbol{x}]\}d\boldsymbol{y}\nonumber \\
= & \rho^{2}r_{\rho}^{2}\int_{D\left(\boldsymbol{0},2r_{\rho}^{-\epsilon}\right)}\{(1-g(\left\Vert \boldsymbol{y}\right\Vert ^{T}))\exp[-\rho r_{\rho}^{2}\int_{A_{\frac{1}{r_{\rho}}}}(g(\left\Vert \boldsymbol{x}\right\Vert ^{T})+g(\left\Vert \boldsymbol{x}-\boldsymbol{y}\right\Vert ^{T})-g(\left\Vert \boldsymbol{x}\right\Vert ^{T})g(\left\Vert \boldsymbol{x}-\boldsymbol{y}\right\Vert ^{T}))d\boldsymbol{x}]\}d\boldsymbol{y}\label{eq:Value-of-b2-finite-rho}
\end{align}
\end{figure*}

Letting $\lambda\triangleq\frac{\log\rho+b}{C}$ for convenience,
noting that (using \eqref{eq:property of toroidal distance 1} and
\eqref{eq:conditions on g(x) - integral boundness})
\[
\lim_{\rho\rightarrow\infty}\int_{A_{\frac{1}{r_{\rho}}}}g(\left\Vert \boldsymbol{x}\right\Vert ^{T})d\boldsymbol{x}=\lim_{\rho\rightarrow\infty}\int_{A_{\frac{1}{r_{\rho}}}}g(\left\Vert \boldsymbol{x}-\boldsymbol{y}\right\Vert ^{T})d\boldsymbol{x}=C
\]
 and that $1-g(\left\Vert \boldsymbol{y}\right\Vert ^{T})\leq1$,
it can further be shown following \eqref{eq:Value-of-b2-finite-rho}
that as $\rho\rightarrow\infty$, 
\begin{align}
 & \lim_{\rho\rightarrow\infty}\lim_{m\rightarrow\infty}\sum_{i\in\Gamma}E(I_{i}^{T}\sum_{j\in\Gamma_{s,i}}I_{j}^{T})\nonumber \\
\leq & e^{-2b}\lim_{\rho\rightarrow\infty}\frac{\lambda}{\rho}\int_{D(\boldsymbol{0},2r_{\rho}^{-\epsilon})}e^{\lambda\int_{A_{\frac{1}{r_{\rho}}}}g(\left\Vert \boldsymbol{x}\right\Vert ^{T})g(\left\Vert \boldsymbol{x}-\boldsymbol{y}\right\Vert ^{T})d\boldsymbol{x}}d\boldsymbol{y}\label{eq:an upper bound on the b2 term}
\end{align}

In the following paragraphs, we will show that the right hand side
of \eqref{eq:an upper bound on the b2 term} converges to $0$ as
$\rho\rightarrow\infty$. Using \eqref{eq:conditions on g(x) - non-increasing}
and \eqref{eq:conditions on g(x) - integral boundness}, we assert
that there exists a positive constant $r$ such that $g\left(r^{-}\right)(1-g\left(r^{+}\right))>0$
where $g\left(r^{-}\right)\triangleq\lim_{x\rightarrow r^{-}}g\left(x\right)$
and $g\left(r^{+}\right)\triangleq\lim_{x\rightarrow r^{+}}g\left(x\right)$.
Indeed if $g$ is a continuous function, any positive constant $r$
with $g\left(r\right)>0$ satisfies the requirement; if $g$ is a
discontinuous function, e.g. a unit disk model, by choosing $r$ to
be the transmission range, $g\left(r^{-}\right)\left(1-g\left(r^{+}\right)\right)=1$. 

In the following discussion we assume that $\rho$ is sufficiently
large such that $\frac{1}{2r_{\rho}}>>2r_{\rho}^{-\epsilon}>>r$.
It can be shown using \eqref{eq:conditions on g(x) - non-increasing},
\eqref{eq:conditions on g(x) - integral boundness} and \eqref{eq:property of toroidal distance 1}
that for $\boldsymbol{y}\in D(\boldsymbol{0},2r_{\rho}^{-\epsilon})$,
\begin{align}
 & \int_{A_{\frac{1}{r_{\rho}}}}g(\left\Vert \boldsymbol{x}\right\Vert ^{T})g(\left\Vert \boldsymbol{x}-\boldsymbol{y}\right\Vert ^{T})d\boldsymbol{x}\nonumber \\
 & \leq\int_{\Re^{2}}g(\left\Vert \boldsymbol{x}\right\Vert )g(\left\Vert \boldsymbol{x}-\boldsymbol{y}\right\Vert )d\boldsymbol{x}\nonumber \\
 & =C-\int_{\Re^{2}}g(\left\Vert \boldsymbol{x}\right\Vert )(1-g(\left\Vert \boldsymbol{x}-\boldsymbol{y}\right\Vert ))d\boldsymbol{x}\nonumber \\
 & \leq C-\int_{D(\boldsymbol{0},r)\backslash D(\boldsymbol{y},r)}g(\left\Vert \boldsymbol{x}\right\Vert )(1-g(\left\Vert \boldsymbol{x}-\boldsymbol{y}\right\Vert ))d\boldsymbol{x}\nonumber \\
 & \leq C-g\left(r^{-}\right)(1-g\left(r^{+}\right))|D(\boldsymbol{0},r)\backslash D(\boldsymbol{y},r)|\label{eq:a bound on the correlation term}
\end{align}
Let $f\left(x\right)\triangleq\pi r^{2}-2r^{2}\arcsin(\sqrt{1-x^{2}/(4r^{2})})+rx\sqrt{1-x^{2}/(4r^{2})}$.
Using some simple geometric analysis, it can be shown that 
\begin{itemize}
\item when $\left\Vert \boldsymbol{y}\right\Vert >2r$, $|D(\boldsymbol{0},r)\backslash D(\boldsymbol{y},r)|=\pi r^{2}$;
and 
\item when $\left\Vert \boldsymbol{y}\right\Vert \leq2r$, $|D(\boldsymbol{0},r)\backslash D(\boldsymbol{y},r)|=f(\left\Vert \boldsymbol{y}\right\Vert )$. 
\end{itemize}
Further, using the definition of $f\left(x\right)$, it can be shown
that 
\begin{itemize}
\item when $\left\Vert \boldsymbol{y}\right\Vert \leq r$, $|D(\boldsymbol{0},r)\backslash D(\boldsymbol{y},r)|\geq\sqrt{3}r\left\Vert \boldsymbol{y}\right\Vert $;
and 
\item when $\left\Vert \boldsymbol{y}\right\Vert >r$, $|D(\boldsymbol{0},r)\backslash D(\boldsymbol{y},r)|\geq(\frac{\pi}{3}+\frac{\sqrt{3}}{2})r^{2}$. 
\end{itemize}
For convenience, let $c_{1}\triangleq g\left(r^{-}\right)(1-g\left(r^{+}\right))\sqrt{3}r$
and $c_{2}\triangleq g\left(r^{-}\right)(1-g\left(r^{+}\right))\left(\frac{\pi}{3}+\frac{\sqrt{3}}{2}\right)r^{2}$.
Noting that $g\left(r^{-}\right)(1-g\left(r^{+}\right))>0$, $c_{1}$
and $c_{2}$ are positive constants, independent of both $\boldsymbol{y}$
and $\rho$. 

As a result of \eqref{eq:a bound on the correlation term} and the
above inequalities on $|D(\boldsymbol{0},r)\backslash D(\boldsymbol{y},r)|$,
it follows that

\begin{align}
 & \lim_{\rho\rightarrow\infty}\frac{\lambda}{\rho}\int_{D(\boldsymbol{0},2r_{\rho}^{-\epsilon})}e^{\lambda\int_{A_{\frac{1}{r_{\rho}}}}g(\left\Vert \boldsymbol{x}\right\Vert ^{T})g(\left\Vert \boldsymbol{x}-\boldsymbol{y}\right\Vert ^{T})d\boldsymbol{x}}d\boldsymbol{y}\nonumber \\
\leq & \lim_{\rho\rightarrow\infty}\frac{\lambda}{\rho}\int_{D\left(\boldsymbol{0},r\right)}e^{\lambda(C-c_{1}\left\Vert \boldsymbol{y}\right\Vert )}d\boldsymbol{y}\nonumber \\
+ & \lim_{\rho\rightarrow\infty}\frac{\lambda}{\rho}\int_{D\left(\boldsymbol{0},2r_{\rho}^{-\epsilon}\right)\backslash D\left(\boldsymbol{0},r\right)}e^{\lambda\left(C-c_{2}\right)}d\boldsymbol{y}\label{eq:decomposition of the integral}
\end{align}
For the first summand in the above equation, it can be shown that:

\begin{align}
 & \lim_{\rho\rightarrow\infty}\frac{\lambda}{\rho}\int_{D\left(\boldsymbol{0},r\right)}e^{\lambda(C-c_{1}\left\Vert \boldsymbol{y}\right\Vert )}d\boldsymbol{y}\nonumber \\
= & \lim_{\rho\rightarrow\infty}\frac{\log\rho+b}{C\rho}\int_{0}^{r}2\pi ye^{\frac{\log\rho+b}{C}(C-c_{1}y)}dy=0\label{eq:first term in the decomposition correlation}
\end{align}

For the second summand in \eqref{eq:decomposition of the integral},
by choosing $\varepsilon<\frac{c_{2}}{C}$ and using \eqref{eq:definition of r_rho},
it follows that 
\begin{align}
 & \lim_{\rho\rightarrow\infty}\frac{\lambda}{\rho}\int_{D(\boldsymbol{0},2r_{\rho}^{-\epsilon})\backslash D\left(\boldsymbol{0},r\right)}e^{\lambda(C-c_{2})}d\boldsymbol{y}\nonumber \\
= & \lim_{\rho\rightarrow\infty}\frac{e^{b(1-\frac{c_{2}}{C})}}{C}\times\frac{\log\rho+b}{\rho^{\frac{c_{2}}{C}}}\times\pi(4r_{\rho}^{-2\epsilon}-r^{2})=0\label{eq:second term in the decomposition correlation}
\end{align}

Combining \eqref{eq:decomposition of the integral}, \eqref{eq:first term in the decomposition correlation}
and \eqref{eq:second term in the decomposition correlation}, it follows
that
\begin{equation}
\lim_{\rho\rightarrow\infty}\frac{\lambda}{\rho}\int_{D(\boldsymbol{0},2r_{\rho}^{-\epsilon})}e^{\lambda\int_{A_{\frac{1}{r_{\rho}}}}g(\left\Vert \boldsymbol{x}\right\Vert ^{T})g(\left\Vert \boldsymbol{x}-\boldsymbol{y}\right\Vert ^{T})d\boldsymbol{x}}d\boldsymbol{y}=0\label{eq:limit of b_2 term}
\end{equation}
As a result of \eqref{eq:an upper bound on the b2 term} and the above
equation: $\lim_{\rho\rightarrow\infty}\lim_{m\rightarrow\infty}b_{2}=0$.

\subsection{An Evaluation of the $b_{3}$ Term\label{sub:An-Evaluation-of-b3}}

We first obtain an analytical expression of the term $E\{I_{i}|(I_{j},j\in\Gamma_{w,i})\}$
in $b_{3}$. Using the same procedure that results in \eqref{eq:ratio correlation},
it can be obtained that (for convenience we use $g_{i}$ for $g(\frac{\left\Vert \boldsymbol{x}-\boldsymbol{x}_{i}\right\Vert ^{T}}{r_{\rho}})$
and use $g_{ij}$ for $g(\frac{\left\Vert \boldsymbol{x}_{i}-\boldsymbol{x}_{j}\right\Vert ^{T}}{r_{\rho}})$
in the following equation): 

\begin{align}
 & \lim_{m\rightarrow\infty}\frac{Pr(I_{i}^{T}=1,I_{j}^{T}=1,I_{k}^{T}=0)}{Pr(I_{i}^{T}=1)Pr(I_{j}^{T}=1,I_{k}^{T}=0)}\nonumber \\
= & \lim_{m\rightarrow\infty}\frac{Pr(I_{i}^{T}=1,I_{j}^{T}=1)-Pr(I_{i}^{T}=1,I_{j}^{T}=1,I_{k}^{T}=1)}{Pr(I_{i}^{T}=1)(Pr(I_{j}^{T}=1)-Pr(I_{j}^{T}=1,I_{k}^{T}=1))}\nonumber \\
\sim_{m} & \left(1-g_{ij}\right)e^{\int_{A}\rho g_{i}g_{j}d\boldsymbol{x}}\nonumber \\
\times & \frac{1-\frac{\rho}{m^{2}}(1-g_{ik})(1-g_{kj})e^{-\int_{A}\rho(g_{k}-g_{i}g_{k}-g_{k}g_{j}+g_{i}g_{j}g_{k})d\boldsymbol{x}}}{1-\frac{\rho}{m^{2}}(1-g_{kj})e^{-\int_{A}\rho(g_{k}-g_{k}g_{j})d\boldsymbol{x}}}\label{eq:correlation of higher order terms}
\end{align}

Using \eqref{eq:conditions on g(x) - non-increasing}, \eqref{eq:conditions on g(x) - integral boundness},
\eqref{eq:scaling property of g(x)} and \eqref{eq:property of toroidal distance 1},
it can be shown that when $j\in\Gamma_{w,i}$ (or equivalently $\left\Vert \boldsymbol{x}_{i}-\boldsymbol{x}_{j}\right\Vert ^{T}>2r_{\rho}^{1-\varepsilon}$),
the integrals of some higher order terms inside the exponential function
in \eqref{eq:correlation of higher order terms} satisfy:

\begin{align*}
 & \int_{A}\rho g(\frac{\left\Vert \boldsymbol{x}-\boldsymbol{x}_{i}\right\Vert ^{T}}{r_{\rho}})g(\frac{\left\Vert \boldsymbol{x}-\boldsymbol{x}_{j}\right\Vert ^{T}}{r_{\rho}})d\boldsymbol{x}\\
= & \int_{D^{T}(\boldsymbol{x}_{i},r_{\rho}^{1-\varepsilon})}\rho g(\frac{\left\Vert \boldsymbol{x}-\boldsymbol{x}_{i}\right\Vert ^{T}}{r_{\rho}})g(\frac{\left\Vert \boldsymbol{x}-\boldsymbol{x}_{j}\right\Vert ^{T}}{r_{\rho}})d\boldsymbol{x}\\
+ & \int_{A\backslash D^{T}(\boldsymbol{x}_{i},r_{\rho}^{1-\varepsilon})}\rho g(\frac{\left\Vert \boldsymbol{x}-\boldsymbol{x}_{i}\right\Vert ^{T}}{r_{\rho}})g(\frac{\left\Vert \boldsymbol{x}-\boldsymbol{x}_{j}\right\Vert ^{T}}{r_{\rho}})d\boldsymbol{x}\\
\leq & 2C\rho r_{\rho}^{2}g(r_{\rho}^{-\varepsilon})\sim_{\rho}o_{\rho}\left(1\right)
\end{align*}
Note also that $g_{ik}=g(\frac{\left\Vert \boldsymbol{x}_{i}-\boldsymbol{x}_{k}\right\Vert ^{T}}{r_{\rho}})=o_{\rho}\left(1\right)$
for $k\in\Gamma_{w,i}$. Using the above equations and \eqref{eq:ratio correlation},
it can be further shown following \eqref{eq:correlation of higher order terms}
that when $j,k\in\Gamma_{w,i}$.
\begin{align}
 & \lim_{\rho\rightarrow\infty}\lim_{m\rightarrow\infty}\frac{Pr(I_{i}^{T}=1,I_{j}^{T}=1,I_{k}^{T}=0)}{Pr(I_{i}^{T}=1)Pr(I_{j}^{T}=1,I_{k}^{T}=0)}\nonumber \\
= & \lim_{\rho\rightarrow\infty}\lim_{m\rightarrow\infty}\frac{Pr(I_{i}^{T}=1,I_{j}^{T}=1)}{Pr(I_{i}^{T}=1)Pr(I_{j}^{T}=1)}\label{eq:a derivation of the correlation terms for b3}
\end{align}
Equation \eqref{eq:a derivation of the correlation terms for b3}
shows that the impact of those events, whose associated indicator
random variables $I_{k}^{T}=0,k\in\Gamma_{w,i}$, on the event $I_{i}^{T}=1$
is asymptotically vanishingly small, hence can be ignored. Denote
by $\Gamma_{i}$ a random set of indices containing all indices $j$
where $j\in\Gamma_{w,i}$ \emph{and }$I_{j}=1$, i.e. the node in
question is also isolated, and denote by $\gamma_{i}$ an instance
of $\Gamma_{i}$. Define $n\triangleq\left|\gamma_{i}\right|$. Following
the same procedure that results in \eqref{eq:a derivation of the correlation terms for b3},
it can be established that (with some verbose but straightforward
discussions omitted) 
\begin{eqnarray}
 &  & \lim_{\rho\rightarrow\infty}\lim_{m\rightarrow\infty}\frac{E\{I_{i}^{T}|(I_{j}^{T},j\in\Gamma_{w,i})\}}{\frac{\rho}{m^{2}}}\nonumber \\
 & = & \lim_{\rho\rightarrow\infty}\lim_{m\rightarrow\infty}\frac{E\{I_{i}^{T}|(I_{j}^{T}=1,j\in\gamma_{i})\}}{\frac{\rho}{m^{2}}}\nonumber \\
 & = & \lim_{\rho\rightarrow\infty}E[e^{-\int_{A}\rho g(\frac{\left\Vert \boldsymbol{x}-\boldsymbol{x}_{i}\right\Vert ^{T}}{r_{\rho}})\prod_{j\in\gamma_{i}}(1-g(\frac{\left\Vert \boldsymbol{x}-\boldsymbol{x}_{j}\right\Vert ^{T}}{r_{\rho}}))d\boldsymbol{x}}\nonumber \\
 & \times & \prod_{j\in\gamma_{i}}(1-g(\frac{\left\Vert \boldsymbol{x}_{i}-\boldsymbol{x}_{j}\right\Vert ^{T}}{r_{\rho}}))]\label{eq:Conditional value I_i b3}
\end{eqnarray}

Equation \eqref{eq:Conditional value I_i b3} gives an analytical
expression of the term $E\{I_{i}^{T}|(I_{j}^{T},j\in\Gamma_{w,i})\}$.
To solve the challenges associated with handling the absolute value
term in $b_{3}$, viz. $|E\{I_{i}^{T}|(I_{j}^{T},j\in\Gamma_{w,i})\}-p_{i}|$,
we further obtain an upper and a lower bound of $I_{i}^{T}|(I_{j}^{T},j\in\Gamma_{w,i})$,
which allows us to remove the absolute value sign in the further analysis
of $b_{3}$.

Note that $\boldsymbol{x}_{i}$ and $\boldsymbol{x}_{j},j\in\Gamma_{w,i}$
is separated by a distance not smaller than $2r_{\rho}^{-\epsilon}$.
Using \eqref{eq:conditions on g(x) - non-increasing}, a lower bound
on the value inside the expectation operator in \eqref{eq:Conditional value I_i b3}
is given by
\begin{eqnarray}
B_{L,i} & \triangleq & (1-g(2r_{\rho}^{-\epsilon}))^{n}e^{-\int_{A}\rho g(\frac{\left\Vert \boldsymbol{x}-\boldsymbol{x}_{i}\right\Vert ^{T}}{r_{\rho}})d\boldsymbol{x}}\label{eq:b3 term lower bound and m2p_i}
\end{eqnarray}
 An upper bound on the value inside the expectation operator in \eqref{eq:Conditional value I_i b3}
is given by
\begin{equation}
B_{U,i}\triangleq e^{-\int_{A}\rho g(\frac{\left\Vert \boldsymbol{x}-\boldsymbol{x}_{i}\right\Vert ^{T}}{r_{\rho}})\prod_{j\in\gamma_{i}}(1-g(\frac{\left\Vert \boldsymbol{x}-\boldsymbol{x}_{j}\right\Vert ^{T}}{r_{\rho}}))d\boldsymbol{x}}\label{eq:definition of B_{u,i}}
\end{equation}
Using $p_{i}=E(I_{i}^{T})$ and \eqref{eq:prob isolated node}, it
can be shown that 
\begin{equation}
B_{U,i}\geq\lim_{m\rightarrow\infty}\frac{m^{2}p_{i}}{\rho}\geq B_{L,i}\label{eq:relation upper bound on b3 and m2p_i}
\end{equation}

Let us consider $E|E\{I_{i}|(I_{j},j\in\Gamma_{w,i})\}-p_{i}|$ now.
From \eqref{eq:Conditional value I_i b3}, \eqref{eq:b3 term lower bound and m2p_i},
\eqref{eq:definition of B_{u,i}} and \eqref{eq:relation upper bound on b3 and m2p_i},
it is clear that 
\begin{eqnarray}
 &  & \lim_{\rho\rightarrow\infty}\lim_{m\rightarrow\infty}\sum_{i\in\Gamma}E|E\{I_{i}^{T}|(I_{j}^{T},j\in\Gamma_{w,i})\}-p_{i}|\nonumber \\
 & \in & [0,\max\{\lim_{\rho\rightarrow\infty}\lim_{m\rightarrow\infty}m^{2}p_{i}-\rho E(B_{L,i}),\nonumber \\
 &  & \lim_{\rho\rightarrow\infty}\lim_{m\rightarrow\infty}\rho E(B_{U,i})-m^{2}p_{i}\}]\label{eq:b3 a bound on the value}
\end{eqnarray}

In the following we will show that both terms $\lim_{m\rightarrow\infty}m^{2}p_{i}-\rho E(B_{L,i})$
and $\lim_{m\rightarrow\infty}\rho E(B_{U,i})-m^{2}p_{i}$ in \eqref{eq:b3 a bound on the value}
approach $0$ as $\rho\rightarrow\infty$. First it can be shown following
\eqref{eq:b3 term lower bound and m2p_i} that

\begin{eqnarray}
 &  & \lim_{m\rightarrow\infty}\rho E(B_{L,i})\nonumber \\
 & \geq & \lim_{m\rightarrow\infty}\rho E[(1-ng(2r_{\rho}^{-\epsilon}))e^{-\int_{A}\rho g(\frac{\left\Vert \boldsymbol{x}-\boldsymbol{x}_{i}\right\Vert ^{T}}{r_{\rho}})d\boldsymbol{x}}]\nonumber \\
 & = & \lim_{m\rightarrow\infty}\rho(1-E\left(n\right)g(2r_{\rho}^{-\epsilon}))e^{-\int_{A}\rho g(\frac{\left\Vert \boldsymbol{x}-\boldsymbol{x}_{i}\right\Vert ^{T}}{r_{\rho}})d\boldsymbol{x}}\label{eq:b3 lower bound x rho finite rho}
\end{eqnarray}
where $\lim_{m\rightarrow\infty}E\left(n\right)$ is the expected
number of isolated nodes in $A\backslash D(\boldsymbol{x}_{i},2r_{\rho}^{1-\epsilon})$.
In the first step of the above equation, the inequality $\left(1-x\right)^{n}\geq1-nx$
for $0\leq x\leq1$ and $n\geq0$ is used. When $\rho\rightarrow\infty$,
$r_{\rho}^{1-\epsilon}\rightarrow0$ and $r_{\rho}^{-\epsilon}\rightarrow\infty$
therefore $\lim_{\rho\rightarrow\infty}\lim_{m\rightarrow\infty}E\left(n\right)=\lim_{\rho\rightarrow\infty}E\left(W^{T}\right)=e^{-b}$
is a bounded value and $\lim_{\rho\rightarrow\infty}\lim_{m\rightarrow\infty}g\left(2r_{\rho}^{-\epsilon}\right)\rightarrow0$,
which is an immediate outcome of \eqref{eq:scaling property of g(x)}.
Using \eqref{eq:value of m2p_i finite rho}, it then follows that
\[
\lim_{\rho\rightarrow\infty}\lim_{m\rightarrow\infty}\frac{\rho E\left(B_{L,i}\right)}{m^{2}p_{i}}\geq\lim_{\rho\rightarrow\infty}\lim_{m\rightarrow\infty}(1-E\left(n\right)g(2r_{\rho}^{-\epsilon}))=1
\]
 Together with \eqref{eq:limiting value m2p_i} and \eqref{eq:relation upper bound on b3 and m2p_i},
we conclude that 
\begin{equation}
\lim_{\rho\rightarrow\infty}\lim_{m\rightarrow\infty}m^{2}p_{i}-\rho E\left(B_{L,i}\right)=0\label{eq:lower bound on b3 equals 0}
\end{equation}
Now let us consider the second term $\lim_{m\rightarrow\infty}\rho E\left(B_{U,i}\right)-m^{2}p_{i}$,
it can be observed that
\begin{align}
 & \lim_{m\rightarrow\infty}E\left(B_{U,i}\right)\nonumber \\
\leq & E[e^{-\int_{D(\boldsymbol{x}_{i},r_{\rho}^{1-\epsilon})}\rho g(\frac{\left\Vert \boldsymbol{x}-\boldsymbol{x}_{i}\right\Vert ^{T}}{r_{\rho}})\prod_{j\in\gamma_{i}}(1-g(\frac{\left\Vert \boldsymbol{x}-\boldsymbol{x}_{j}\right\Vert ^{T}}{r_{\rho}}))d\boldsymbol{x}}]\nonumber \\
\leq & \lim_{m\rightarrow\infty}E[e^{-\int_{D(\boldsymbol{x}_{i},r_{\rho}^{1-\epsilon})}\rho g(\frac{\left\Vert \boldsymbol{x}-\boldsymbol{x}_{i}\right\Vert ^{T}}{r_{\rho}})\prod_{j\in\gamma_{i}}(1-g(\frac{r_{\rho}^{1-\epsilon}}{r_{\rho}}))d\boldsymbol{x}}]\nonumber \\
= & \lim_{m\rightarrow\infty}E(e^{-(1-g(r_{\rho}^{-\epsilon}))^{n}\int_{D(\boldsymbol{x}_{i},r_{\rho}^{1-\epsilon})}\rho g(\frac{\left\Vert \boldsymbol{x}-\boldsymbol{x}_{i}\right\Vert ^{T}}{r_{\rho}})d\boldsymbol{x}})\nonumber \\
\leq & \lim_{m\rightarrow\infty}E(e^{-(1-ng(r_{\rho}^{-\epsilon}))\int_{D(\boldsymbol{x}_{i},r_{\rho}^{1-\epsilon})}\rho g(\frac{\left\Vert \boldsymbol{x}-\boldsymbol{x}_{i}\right\Vert ^{T}}{r_{\rho}})d\boldsymbol{x}})\label{eq:limiting value of the upper bound final step}
\end{align}
where in the second step, the non-increasing property of $g$, and
the fact that $\boldsymbol{x}_{j}$ is located in $A\backslash D(\boldsymbol{x}_{i},2r_{\rho}^{1-\epsilon})$
and $\boldsymbol{x}$ is located in $D(\boldsymbol{x}_{i},r_{\rho}^{1-\epsilon})$,
therefore $\left\Vert \boldsymbol{x}-\boldsymbol{x}_{j}\right\Vert ^{T}\geq r_{\rho}^{1-\epsilon}$
is used. It can be further demonstrated that the term $\int_{D(\boldsymbol{x}_{i},r_{\rho}^{1-\epsilon})}\rho g(\frac{\left\Vert \boldsymbol{x}-\boldsymbol{x}_{i}\right\Vert }{r_{\rho}})d\boldsymbol{x}$
in \eqref{eq:limiting value of the upper bound final step} have the
following property: 
\begin{eqnarray}
\eta\left(\varepsilon,\rho\right) & \triangleq & \int_{D(\boldsymbol{x}_{i},r_{\rho}^{1-\epsilon})}\rho g(\frac{\left\Vert \boldsymbol{x}-\boldsymbol{x}_{i}\right\Vert ^{T}}{r_{\rho}})d\boldsymbol{x}\nonumber \\
 & = & \rho r_{\rho}^{2}\int_{D(\frac{\boldsymbol{x}_{i}}{r_{\rho}},r_{\rho}^{-\epsilon})}g(\left\Vert \boldsymbol{x}-\boldsymbol{x}_{i}/r_{\rho}\right\Vert ^{T})d\boldsymbol{x}\nonumber \\
 & \leq & C\rho r_{\rho}^{2}=\log\rho+b\label{eq:b3 upper bound the integral}
\end{eqnarray}
For the other term $ng(r_{\rho}^{-\epsilon})$ in \eqref{eq:limiting value of the upper bound final step},
choosing a positive constant $\delta<2\epsilon$ and using Markov's
inequality, it can be shown that $Pr(n\geq r_{\rho}^{-\delta})\leqslant r_{\rho}^{\delta}E\left(n\right)$.
Therefore
\begin{eqnarray*}
 &  & \lim_{\rho\rightarrow\infty}\lim_{m\rightarrow\infty}Pr(ng(r_{\rho}^{-\epsilon})\eta(\varepsilon,\rho)\geq r_{\rho}^{-\delta}g(r_{\rho}^{-\epsilon})\eta(\varepsilon,\rho))\\
 & \leq & \lim_{\rho\rightarrow\infty}\lim_{m\rightarrow\infty}r_{\rho}^{\delta}E\left(n\right)
\end{eqnarray*}
where $\lim_{\rho\rightarrow\infty}r_{\rho}^{-\delta}g(r_{\rho}^{-\epsilon})\eta(\varepsilon,\rho)=0$
due to \eqref{eq:scaling property of g(x)}, \eqref{eq:b3 upper bound the integral}
and $\delta<2\epsilon$, $\lim_{\rho\rightarrow\infty}r_{\rho}^{B}=0$
for any positive constant $B$, and $\lim_{\rho\rightarrow\infty}\lim_{m\rightarrow\infty}r_{\rho}^{\delta}E\left(n\right)=0$
due to that $\lim_{\rho\rightarrow\infty}\lim_{m\rightarrow\infty}E\left(n\right)=\lim_{\rho\rightarrow\infty}E(W^{T})=e^{-b}$
is a bounded value and that $\lim_{\rho\rightarrow\infty}r_{\rho}^{\delta}=0$.
Therefore
\begin{equation}
\lim_{\rho\rightarrow\infty}\lim_{m\rightarrow\infty}Pr(ng(r_{\rho}^{-\epsilon})\eta(\varepsilon,\rho)=0)=1\label{eq:asymptotic almost surely expected n}
\end{equation}

As a result of \eqref{eq:property of toroidal distance 1}, \eqref{eq:limiting value of the upper bound final step},
\eqref{eq:b3 upper bound the integral} and \eqref{eq:asymptotic almost surely expected n}:
\begin{eqnarray}
 &  & \lim_{\rho\rightarrow\infty}\lim_{m\rightarrow\infty}\rho E(B_{U,i})\nonumber \\
 & \leq & \lim_{\rho\rightarrow\infty}\lim_{m\rightarrow\infty}\rho E(e^{-\int_{D(\boldsymbol{x}_{i},r_{\rho}^{1-\epsilon})}\rho g(\frac{\left\Vert \boldsymbol{x}-\boldsymbol{x}_{i}\right\Vert ^{T}}{r_{\rho}})d\boldsymbol{x}})\nonumber \\
 & = & \lim_{\rho\rightarrow\infty}\rho e^{-\int_{D(\boldsymbol{0},r_{\rho}^{1-\varepsilon})}\rho g(\frac{\left\Vert \boldsymbol{x}\right\Vert }{r_{\rho}})d\boldsymbol{x}}\nonumber \\
 & = & \lim_{\rho\rightarrow\infty}\rho e^{-\rho r_{\rho}^{2}(C-\int_{\Re^{2}\backslash D(\boldsymbol{0},r_{\rho}^{-\varepsilon})}g\left(\left\Vert \boldsymbol{x}\right\Vert \right)d\boldsymbol{x})}\nonumber \\
 & = & e^{-b}\lim_{\rho\rightarrow\infty}e^{\rho r_{\rho}^{2}\int_{\Re^{2}\backslash D(\boldsymbol{0},r_{\rho}^{-\varepsilon})}g\left(\left\Vert \boldsymbol{x}\right\Vert \right)d\boldsymbol{x}}=e^{-b}\label{eq:limiting value of B_U}
\end{eqnarray}
where the last step results because

\begin{align}
 & \lim_{\rho\rightarrow\infty}\rho r_{\rho}^{2}\int_{\Re^{2}\backslash D(\boldsymbol{0},r_{\rho}^{-\varepsilon})}g\left(\left\Vert \boldsymbol{x}\right\Vert \right)d\boldsymbol{x}\nonumber \\
= & \lim_{\rho\rightarrow\infty}\frac{\int_{r_{\rho}^{-\varepsilon}}^{\infty}2\pi xg\left(x\right)dx}{\frac{C}{\log\rho+b}}\nonumber \\
= & \lim_{\rho\rightarrow\infty}\frac{\pi\varepsilon r_{\rho}^{-\varepsilon}g(r_{\rho}^{-\varepsilon})r_{\rho}^{-\varepsilon-2}\frac{\log\rho+b-1}{C\rho^{2}}}{\frac{C}{\rho(\log\rho+b)^{2}}}\label{eq:little's rule trunction}\\
= & \lim_{\rho\rightarrow\infty}\frac{\pi\varepsilon}{C}(\log\rho+b)^{2}r_{\rho}^{-2\varepsilon}o_{\rho}(\frac{1}{r_{\rho}^{-2\varepsilon}\log^{2}(r_{\rho}^{-2\varepsilon})})=0\label{eq:diminishing truncation effect}
\end{align}
where L'H�pital's rule is used in reaching \eqref{eq:little's rule trunction}
and in the third step \eqref{eq:Condition on g(x) requirement 2}
is used. Using \eqref{eq:limiting value m2p_i}, \eqref{eq:relation upper bound on b3 and m2p_i}
and \eqref{eq:limiting value of B_U}, it can be shown that 
\begin{equation}
\lim_{\rho\rightarrow\infty}\lim_{m\rightarrow\infty}\rho E(B_{U,i})-m^{2}p_{i}=0\label{eq:upper bound on b3 equals 0}
\end{equation}

As a result of \eqref{eq:b3 a bound on the value}, \eqref{eq:lower bound on b3 equals 0}
and \eqref{eq:upper bound on b3 equals 0}, $\lim_{\rho\rightarrow\infty}\lim_{m\rightarrow\infty}b_{3}=0$.

A combination of the analysis in subsections A, B and C completes
this proof.

\section*{Appendix II: Proof of Theorem \ref{thm:component of unbounded order}}

For notational convenience, we prove the result for $\mathcal{G}\left(\mathcal{X}_{\lambda},g,A_{\frac{1}{r_{\rho}}}\right)$
and the result is equally valid for $\mathcal{G}\left(\mathcal{X}_{\rho},g_{r_{\rho}},A\right)$.
The proof is based on analyzing the number of components in $\mathcal{G}\left(\mathcal{X}_{\lambda},g,A_{\frac{1}{r_{\rho}}}\right)$
of order greater than some integer $M$ as $\rho\rightarrow\infty$.
Specifically we will show that $\lim_{M\rightarrow\infty}\lim_{\rho\rightarrow\infty}\Pr(\xi_{>M}=1)=1$.

A direct analysis of $\Pr(\xi_{>M}=1)$ can be difficult. In this
paper, we first analyze $E(\xi_{>M})$ and then use the result on
$E(\xi_{>M})$ to establish the desired asymptotic result on $\Pr(\xi_{>M}=1)$.

Denote by $g_{1}\left(\boldsymbol{x}_{1},\ldots,\boldsymbol{x}_{k}\right)$
the probability that a set of $k$ nodes at non-random positions $\boldsymbol{x}_{1}$,
$\ldots$, $\boldsymbol{x}_{k}\in A_{\frac{1}{r_{\rho}}}$ forms a
connected component where nodes are connected randomly and independently
following the connection function $g$. Denote by $g_{2}\left(\boldsymbol{y};\boldsymbol{x}_{1},\boldsymbol{x}_{2},\ldots,\boldsymbol{x}_{k}\right)$
the probability that a node at non-random position $\boldsymbol{y}$
is connected to at least one node in $\left\{ \boldsymbol{x}_{1},\boldsymbol{x}_{2},\ldots,\boldsymbol{x}_{k}\right\} $.
As an easy consequence of \cite[Lemma 4]{Mao11Towards}, which showed
that the expected number of components of order $k$, denoted by $\xi_{k}$
, in $\mathcal{G}\left(\mathcal{X}_{\lambda},g,A_{\frac{1}{r_{\rho}}}\right)$
is given by $E\left(\xi_{k}\right)=\frac{\lambda^{k}}{k!}\int_{(A_{\frac{1}{r_{\rho}}})^{k}}g_{1}(\boldsymbol{x}_{1},\ldots,\boldsymbol{x}_{k})e^{-\lambda\int_{A_{\frac{1}{r_{\rho}}}}g_{2}(\boldsymbol{y};\boldsymbol{x}_{1},\ldots,\boldsymbol{x}_{k})d\boldsymbol{y}}d(\boldsymbol{x}_{1}\cdots\boldsymbol{x}_{k})$,
it follows that
\begin{align}
 & E\left(\xi_{>M}\right)\nonumber \\
= & \sum_{k=M+1}^{\infty}\frac{\lambda^{k}}{k!}\int_{(A_{\frac{1}{r_{\rho}}})^{k}}(g_{1}(\boldsymbol{x}_{1},\ldots,\boldsymbol{x}_{k})\nonumber \\
 & e^{-\lambda\int_{A_{\frac{1}{r_{\rho}}}}g_{2}(\boldsymbol{y};\boldsymbol{x}_{1},\ldots,\boldsymbol{x}_{k})d\boldsymbol{y}})d(\boldsymbol{x}_{1}\cdots\boldsymbol{x}_{k})\nonumber \\
\leq & \sum_{k=M+1}^{\infty}\frac{\lambda^{k}}{k!}\int_{(A_{\frac{1}{r_{\rho}}})^{k}}e^{-\lambda\int_{A_{\frac{1}{r_{\rho}}}}g_{2}(\boldsymbol{y};\boldsymbol{x}_{1},\ldots,\boldsymbol{x}_{k})d\boldsymbol{y}}d(\boldsymbol{x}_{1}\cdots\boldsymbol{x}_{k})\nonumber \\
= & \sum_{k=1}^{\infty}\frac{\lambda^{k}}{k!}\int_{(A_{\frac{1}{r_{\rho}}})^{k}}e^{-\lambda\int_{A_{\frac{1}{r_{\rho}}}}g_{2}(\boldsymbol{y};\boldsymbol{x}_{1},\ldots,\boldsymbol{x}_{k})d\boldsymbol{y}}d(\boldsymbol{x}_{1}\cdots\boldsymbol{x}_{k})\nonumber \\
- & \sum_{k=1}^{M}\frac{\lambda^{k}}{k!}\int_{(A_{\frac{1}{r_{\rho}}})^{k}}e^{-\lambda\int_{A_{\frac{1}{r_{\rho}}}}g_{2}(\boldsymbol{y};\boldsymbol{x}_{1},\ldots,\boldsymbol{x}_{k})d\boldsymbol{y}}d(\boldsymbol{x}_{1}\cdots\boldsymbol{x}_{k})\label{eq:number of components of order greater than M - nonasymptotic}
\end{align}

In the following we show that as $\rho\rightarrow\infty$, the first
term in \eqref{eq:number of components of order greater than M - nonasymptotic}
converges to $e^{e^{-b}}$, and the second term in \eqref{eq:number of components of order greater than M - nonasymptotic}
after the ``$-$'' sign is lower-bounded by $\sum_{k=1}^{M}\frac{(e^{-b})^{k}}{k!}$.
The conclusion then follows that $E(\xi_{>M})$ converge to $1$ as
$\rho\rightarrow\infty$ and $M\rightarrow\infty$.

Let us consider the first term in \eqref{eq:number of components of order greater than M - nonasymptotic}
now. Let 
\begin{equation}
\Phi\triangleq\lambda\int_{A_{\frac{1}{r_{\rho}}}}[1-g_{2}(\boldsymbol{y};\boldsymbol{x}_{1},\ldots\boldsymbol{x}_{k})]d\boldsymbol{y}\label{eq:definition of fai}
\end{equation}
 for convenience. It can be shown that 
\begin{align}
 & \lim_{\rho\rightarrow\infty}\sum_{k=1}^{\infty}\frac{\lambda^{k}}{k!}\int_{(A_{\frac{1}{r_{\rho}}})^{k}}e^{-\lambda\int_{A_{\frac{1}{r_{\rho}}}}g_{2}(\boldsymbol{y};\boldsymbol{x}_{1},\ldots\boldsymbol{x}_{k})d\boldsymbol{y}}d(\boldsymbol{x}_{1},\ldots\boldsymbol{x}_{k})\nonumber \\
= & \lim_{\rho\rightarrow\infty}\sum_{k=1}^{\infty}\frac{\lambda^{k}}{k!}e^{-\rho}\int_{(A_{\frac{1}{r_{\rho}}})^{k}}e^{\Phi}d(\boldsymbol{x}_{1},\ldots\boldsymbol{x}_{k})\nonumber \\
= & \lim_{\rho\rightarrow\infty}\sum_{k=1}^{\infty}\frac{\lambda^{k}}{k!}e^{-\rho}\int_{(A_{\frac{1}{r_{\rho}}})^{k}}\sum_{n=0}^{\infty}\frac{\Phi^{n}}{n!}d(\boldsymbol{x}_{1},\ldots\boldsymbol{x}_{k})\nonumber \\
= & \sum_{n=0}^{\infty}\frac{1}{n!}\lim_{\rho\rightarrow\infty}\sum_{k=1}^{\infty}\frac{\lambda^{k}}{k!}e^{-\rho}\int_{(A_{\frac{1}{r_{\rho}}})^{k}}\Phi^{n}d(\boldsymbol{x}_{1},\ldots\boldsymbol{x}_{k})\label{eq:an intermediate result on the first term in components>M}
\end{align}

Next we shall show that in \eqref{eq:an intermediate result on the first term in components>M},
$\lim_{\rho\rightarrow\infty}\sum_{k=1}^{\infty}\frac{\lambda^{k}}{k!}e^{-\rho}\int_{(A_{\frac{1}{r_{\rho}}})^{k}}\Phi^{n}d(\boldsymbol{x}_{1},\ldots\boldsymbol{x}_{k})=(e^{-b})^{n}$.
Given this result, conclusion readily follows from \eqref{eq:an intermediate result on the first term in components>M}
that the first term in \eqref{eq:number of components of order greater than M - nonasymptotic}
converges to $e^{e^{-b}}$. 

A direct computation of the term $\lim_{\rho\rightarrow\infty}\sum_{k=1}^{\infty}\frac{\lambda^{k}}{k!}e^{-\rho}\int_{(A_{\frac{1}{r_{\rho}}})^{k}}\Phi^{n}d(\boldsymbol{x}_{1},\ldots\boldsymbol{x}_{k})$
turns out to be very difficult. To resolve the difficulty, we construct
a random integer $X$, depending on $\rho$, such that on the one
hand, the pmf (probability mass function) of $X$ has an analytical
form that can be easily related to the term $\sum_{k=1}^{\infty}\frac{\lambda^{k}}{k!}e^{-\rho}\int_{(A_{\frac{1}{r_{\rho}}})^{k}}\Phi^{n}d(\boldsymbol{x}_{1},\ldots\boldsymbol{x}_{k})$;
and on the other hand using the Chen-Stein bound we are familiar with,
the pmf can be shown to converge to a Poisson distribution as $\rho\rightarrow\infty$.
In this way, we are able to compute the above term using the intermediate
random integer $X$. In the following, we give details of the analysis.

We first construct the random integer $X$ described in the last paragraph
and demonstrate its properties related to our analysis.

Consider an additional \emph{independent} Poisson point process $\mathcal{X}'_{\lambda}$
with nodes Poissonly distributed on $A_{\frac{1}{r_{\rho}}}$ and
with density $\lambda$, being added to $\mathcal{G}\left(\mathcal{X}_{\lambda},g,A_{\frac{1}{r_{\rho}}}\right)$.
Further, nodes in $\mathcal{X}'_{\lambda}$ are connected with nodes
in $\mathcal{X}_{\lambda}$ following $g$ independently, i.e. a node
in $\mathcal{X}'_{\lambda}$ and a node in $\mathcal{X}_{\lambda}$
separated by an Euclidean distance $x$ are connected with probability
$g\left(x\right)$, independent of any other connection.

Let $X$ be the number of nodes in $\mathcal{X}'_{\lambda}$ that
are \emph{not }directly connected to any node in $\mathcal{X}_{\lambda}$.
It is evident that, conditioned on $\mathcal{X}_{\lambda}=(\boldsymbol{x}_{1},\ldots\boldsymbol{x}_{k})$
where $\boldsymbol{x}_{1},\ldots\boldsymbol{x}_{k}\in A_{\frac{1}{r_{\rho}}}$
and $|\mathcal{X}_{\lambda}|>0$, a randomly chosen node in $\mathcal{X}'_{\lambda}$
at location $\boldsymbol{y}$ is \emph{not }directly connected to
any node in $\mathcal{X}_{\lambda}$ with probability $1-g_{2}(\boldsymbol{y};\boldsymbol{x}_{1},\ldots,\boldsymbol{x}_{k})$
, which is determined by its location only. It readily follows that
the conditional distribution of $X$, i.e. $X|\mathcal{X}_{\lambda}=(\boldsymbol{x}_{1},\ldots\boldsymbol{x}_{k})$,
is Poisson with mean $\lambda\int_{A_{\frac{1}{r_{\rho}}}}[1-g_{2}(\boldsymbol{y};\boldsymbol{x}_{1},\ldots,\boldsymbol{x}_{k})]d\boldsymbol{y}$
\cite{Meester96Continuum}. As a result of the above discussion:

\begin{equation}
\Pr(X=m|\mathcal{X}_{\lambda}=(\boldsymbol{x}_{1},\ldots\boldsymbol{x}_{k}))=\frac{\Phi^{m}}{m!}e^{-\Phi}\label{eq:conditional probability x=00003Dm-1}
\end{equation}

Obviously when $\mathcal{X}_{\lambda}=\emptyset$, $\Pr(X=m|\mathcal{X}_{\lambda}=\emptyset)=\Pr(|\mathcal{X}'_{\lambda}|=m)$.
Therefore the unconditional distribution of $X$ is given by:
\begin{align}
 & \Pr\left(X=m\right)\nonumber \\
= & \sum_{k=1}^{\infty}\frac{\lambda^{k}}{k!}e^{-\rho}\int_{(A_{\frac{1}{r_{\rho}}})^{k}}\frac{\Phi^{m}}{m!}e^{-\Phi}d(\boldsymbol{x}_{1},\ldots\boldsymbol{x}_{k})+\frac{\rho^{m}}{m!}e^{-2\rho}\label{eq:unconditional probability xm k0}
\end{align}

Note that as $\rho\rightarrow\infty$, the term $\frac{\rho^{m}}{m!}e^{-2\rho}$
in \eqref{eq:unconditional probability xm k0}, which is associated
with $\mathcal{X}_{\lambda}=\emptyset$, becomes vanishingly small.
Further note that $\sum_{m=0}^{\infty}\frac{\rho^{m}}{m!}e^{-2\rho}=e^{-\rho}\rightarrow0$
as $\rho\rightarrow\infty$, i.e. as $\rho\rightarrow\infty$ even
the cumulative contribution to the cdf of $X$ is negligibly small. 

If we define $g_{2}\left(\boldsymbol{y};\emptyset\right)\triangleq0$
for completeness, we can also write \eqref{eq:unconditional probability xm k0}
as
\begin{equation}
\Pr\left(X=m\right)=\sum_{k=0}^{\infty}\frac{\lambda^{k}}{k!}e^{-\rho}\int_{(A_{\frac{1}{r_{\rho}}})^{k}}\frac{\Phi^{m}}{m!}e^{-\Phi}d(\boldsymbol{x}_{1},\ldots\boldsymbol{x}_{k})\label{eq:unconditional probability x m}
\end{equation}

Using \eqref{eq:unconditional probability x m}, it can be readily
shown that
\begin{align}
 & E\left(X\right)=\sum_{m=0}^{\infty}m\Pr\left(X=m\right)\nonumber \\
= & \sum_{k=0}^{\infty}\frac{\lambda^{k}}{k!}e^{-\rho}\int_{(A_{\frac{1}{r_{\rho}}})^{k}}\Phi d(\boldsymbol{x}_{1},\ldots\boldsymbol{x}_{k})\nonumber \\
= & \sum_{k=0}^{\infty}\frac{\lambda^{k}}{k!}e^{-\rho}\{\lambda\int_{A_{\frac{1}{r_{\rho}}}}\{\int_{A_{\frac{1}{r_{\rho}}}}[1-g(\left\Vert \boldsymbol{x}-\boldsymbol{y}\right\Vert )]d\boldsymbol{x}\}^{k}d\boldsymbol{y}\}\nonumber \\
= & \lambda\int_{A_{\frac{1}{r_{\rho}}}}e^{-\lambda\int_{A_{\frac{1}{r_{\rho}}}}g(\left\Vert \boldsymbol{x}-\boldsymbol{y}\right\Vert )d\boldsymbol{x}}d\boldsymbol{y}\label{eq:expected number of disconnected nodes-1}
\end{align}

Comparing the above equation with \cite[Theorem 1]{Mao11Towards},
the conclusion readily follows that the above value is equal to the
expected number of isolated nodes in $\mathcal{G}\left(\mathcal{X}_{\lambda},g,A_{\frac{1}{r_{\rho}}}\right)$,
denoted by $W$. It then follows from \cite[Theorem 1]{Mao11Towards},
that $\lim_{\rho\rightarrow\infty}E\left(X\right)=e^{-b}$. In fact
a stronger result that the distributions of $X$ and $W$ converge
to the same Poisson distribution as $\rho\rightarrow\infty$ can be
established:
\begin{lem}
\label{lem:convergence of X to Poisson}As $\rho\rightarrow\infty$,
the distribution of $X$ converges to a Poisson distribution with
mean $e^{-b}$, i.e. the total variation distance between the distribution
of $X$ and a Poisson distribution with mean $e^{-b}$ reduces to
$0$ as $\rho\rightarrow\infty$.
\end{lem}
Lemma \ref{lem:convergence of X to Poisson} can be proved using exactly
the same steps as those used in proving Theorem \ref{thm:Number of isolated nodes square}.
Therefore the proof is omitted.

As a result of Lemma \ref{lem:convergence of X to Poisson}, for an
arbitrary set of non-negative integers, denoted by $\Gamma$,
\begin{equation}
\lim_{\rho\rightarrow\infty}\sum_{m\in\Gamma}\Pr(X=m)=\sum_{m\in\Gamma}\frac{(e^{-b})^{m}}{m!}e^{-e^{-b}}\label{eq:asymptotic unconditional probability X m}
\end{equation}

Now we are ready to continue our analysis on $\lim_{\rho\rightarrow\infty}\sum_{k=1}^{\infty}\frac{\lambda^{k}}{k!}e^{-\rho}\int_{(A_{\frac{1}{r_{\rho}}})^{k}}\Phi^{n}d(\boldsymbol{x}_{1},\ldots\boldsymbol{x}_{k})$.
Using \eqref{eq:unconditional probability xm k0} first and then using
\eqref{eq:asymptotic unconditional probability X m}, it can be shown
that for any positive integer $n$:

\begin{align*}
 & \lim_{\rho\rightarrow\infty}\sum_{k=1}^{\infty}\frac{\lambda^{k}}{k!}e^{-\rho}\int_{(A_{\frac{1}{r_{\rho}}})^{k}}\Phi^{n}d(\boldsymbol{x}_{1},\ldots\boldsymbol{x}_{k})\\
= & \lim_{\rho\rightarrow\infty}\sum_{k=1}^{\infty}\frac{\lambda^{k}}{k!}e^{-\rho}\int_{(A_{\frac{1}{r_{\rho}}})^{k}}\sum_{m=0}^{\infty}\Phi^{n}\frac{\Phi^{m}}{m!}e^{-\Phi}d(\boldsymbol{x}_{1},\ldots\boldsymbol{x}_{k})\\
= & \sum_{m=0}^{\infty}\frac{1}{m!}\lim_{\rho\rightarrow\infty}\sum_{k=1}^{\infty}\frac{\lambda^{k}}{k!}e^{-\rho}\int_{(A_{\frac{1}{r_{\rho}}})^{k}}\Phi^{n+m}e^{-\Phi}d(\boldsymbol{x}_{1},\ldots\boldsymbol{x}_{k})\\
= & \sum_{m=0}^{\infty}\frac{1}{m!}\lim_{\rho\rightarrow\infty}(\Pr\left(X=n+m\right)-\frac{\rho^{(n+m)}}{(n+m)!}e^{-2\rho})(n+m)!\\
= & \sum_{m=0}^{\infty}\frac{(e^{-b})^{n+m}}{m!}e^{-e^{-b}}=(e^{-b})^{n}
\end{align*}
Using the above equation, it follows from \eqref{eq:an intermediate result on the first term in components>M}
that

\begin{align}
 & \lim_{\rho\rightarrow\infty}\sum_{k=1}^{\infty}\frac{\lambda^{k}}{k!}\int_{(A_{\frac{1}{r_{\rho}}})^{k}}e^{-\lambda\int_{A_{\frac{1}{r_{\rho}}}}g_{2}(\boldsymbol{y};\boldsymbol{x}_{1},\ldots\boldsymbol{x}_{k})d\boldsymbol{y}}d(\boldsymbol{x}_{1},\ldots\boldsymbol{x}_{k})\nonumber \\
= & \sum_{n=0}^{\infty}\frac{(e^{-b})^{n}}{n!}=e^{e^{-b}}\label{eq:asymptotic total number of components}
\end{align}

This deals with the first term on the right of \eqref{eq:number of components of order greater than M - nonasymptotic}.
Now we continue with the analysis of the second term in \eqref{eq:number of components of order greater than M - nonasymptotic}.
As an easy consequence of the union bound, $g_{2}\left(\boldsymbol{y};\boldsymbol{x}_{1},\boldsymbol{x}_{2},\ldots,\boldsymbol{x}_{k}\right)\leq\sum_{i=1}^{k}g\left(\left\Vert \boldsymbol{y}-\boldsymbol{x}_{i}\right\Vert \right)$,
it can then be shown that
\begin{eqnarray}
 &  & \frac{\lambda^{k}}{k!}\int_{(A_{\frac{1}{r_{\rho}}})^{k}}e^{-\lambda\int_{A_{\frac{1}{r_{\rho}}}}g_{2}(\boldsymbol{y};\boldsymbol{x}_{1},\ldots\boldsymbol{x}_{k})d\boldsymbol{y}}d(\boldsymbol{x}_{1}\cdots\boldsymbol{x}_{k})\nonumber \\
 & \geq & \frac{\lambda^{k}}{k!}\int_{(A_{\frac{1}{r_{\rho}}})^{k}}e^{-\lambda\int_{A_{\frac{1}{r_{\rho}}}}\sum_{i=1}^{k}g(\left\Vert \boldsymbol{y}-\boldsymbol{x}_{i}\right\Vert )d\boldsymbol{y}}d(\boldsymbol{x}_{1}\cdots\boldsymbol{x}_{k})\nonumber \\
 & = & \frac{(\lambda\int_{A_{\frac{1}{r_{\rho}}}}e^{-\lambda\int_{A_{\frac{1}{r_{\rho}}}}g(\left\Vert \boldsymbol{x}-\boldsymbol{y}\right\Vert )d\boldsymbol{y}}d\boldsymbol{x})^{k}}{k!}\label{eq:a bound on the number of components of order k}
\end{eqnarray}
and using \cite[Theorem 1]{Mao11Towards}, it can be further shown
that

\begin{align}
 & \lim_{\rho\rightarrow\infty}\frac{\lambda^{k}}{k!}\int_{(A_{\frac{1}{r_{\rho}}})^{k}}e^{-\lambda\int_{A_{\frac{1}{r_{\rho}}}}g_{2}(\boldsymbol{y};\boldsymbol{x}_{1},\ldots\boldsymbol{x}_{k})d\boldsymbol{y}}d(\boldsymbol{x}_{1}\cdots\boldsymbol{x}_{k})\nonumber \\
\geq & \frac{(e^{-b})^{k}}{k!}\label{eq:asymptotic bound on the number of component of order k}
\end{align}

Note that \eqref{eq:asymptotic bound on the number of component of order k}
can also be obtained from Jensen's inequality.

Combining \eqref{eq:number of components of order greater than M - nonasymptotic},
\eqref{eq:asymptotic total number of components} and \eqref{eq:asymptotic bound on the number of component of order k},
it follows that
\begin{equation}
\lim_{\rho\rightarrow\infty}E(\xi_{>M})\leq e^{e^{-b}}-\sum_{k=1}^{M}\frac{(e^{-b})^{k}}{k!}=1+\frac{(\eta_{M})^{M+1}}{(M+1)!}\label{eq:asymptotic bound on the number of components of order >M}
\end{equation}
where in the last step Taylor's theorem is used, $\eta_{M}$ is a
number depending on $M$ and $ $$0\leq\eta_{M}\leq e^{-b}$.

In Theorems \ref{thm:Number of isolated nodes square} and \cite[Theorem 4]{Mao11Towards},
we have established respectively that the asymptotic distribution
of the number of isolated nodes in $\mathcal{G}\left(\mathcal{X}_{\lambda},g,A_{\frac{1}{r_{\rho}}}\right)$
is Poisson with mean $e^{-b}$ and the number of components in $\mathcal{G}\left(\mathcal{X}_{\lambda},g,A_{\frac{1}{r_{\rho}}}\right)$
of order within $\left[2,M\right]$ vanishes as $\rho\rightarrow\infty$.
As a consequence of the above two results, 
\begin{equation}
\lim_{\rho\rightarrow\infty}\Pr(\xi_{>M}\geq1)=1\;\;\textrm{and}\;\;\lim_{\rho\rightarrow\infty}\Pr(\xi_{>M}=0)=0\label{eq:existsnce of component of giant component}
\end{equation}
Further note that
\begin{align}
 & E(\xi_{>M})=\sum_{m=1}^{\infty}m\Pr(\xi_{>M}=m)\nonumber \\
\geq & \Pr(\xi_{>M}=1)+2\sum_{m=2}^{\infty}\Pr(\xi_{>M}=m)\nonumber \\
= & \Pr(\xi_{>M}=1)+2(1-\Pr(\xi_{>M}=1)-\Pr(\xi_{>M}=0))\label{eq:inequality on the number of component>M}
\end{align}

Combing the three equations \eqref{eq:asymptotic bound on the number of components of order >M},
\eqref{eq:existsnce of component of giant component} and \eqref{eq:inequality on the number of component>M}:
\begin{equation}
\lim_{\rho\rightarrow\infty}\Pr(\xi_{>M}=1)\geq1-\frac{(\eta_{M})^{M+1}}{\left(M+1\right)!}\label{eq:convergence of number of components of order greater than M}
\end{equation}

As an easy consequence of the above equation: 
\[
\lim_{M\rightarrow\infty}\lim_{\rho\rightarrow\infty}\Pr\left(\xi_{>M}=1\right)=1
\]

\bibliographystyle{IEEEtran}

\begin{IEEEbiography}
{Guoqiang Mao} (S'98\textendash{}M'02\textendash{}SM\textquoteright{}08)
received PhD in telecommunications engineering in 2002 from Edith
Cowan University, Australia. He joined the School of Electrical and
Information Engineering, the University of Sydney in December 2002.
He has published over 100 papers in international journals and conferences.
His research interests include wireless multihop networks (e.g. vehicular
networks, mesh networks mobile networks, delay-tolerant networks,
opportunistic networks), wireless sensor networks, wireless localization
techniques, applied graph theory and network performance analysis.
He is a Senior Member of IEEE and an Associate Editor of IEEE Transactions
on Vehicular Technology. He has served as a program committee member
in a large number of international conferences. He was a symposium
co-chair of IEEE PIMRC 2012, a publicity co-chair of 2007 SenSys and
2010 IEEE WCNC.
\end{IEEEbiography}

\begin{IEEEbiography}
{Brian D.O. Anderson} (S\textquoteright{}62\textendash{}M\textquoteright{}66\textendash{}SM\textquoteright{}74\textendash{}F\textquoteright{}75\textendash{}
LF\textquoteright{}07) was born in Sydney, Australia, and educated
at Sydney University in mathematics and electrical engineering, with
PhD in electrical engineering from Stanford University in 1966. He
is a Distinguished Professor at the Australian National University
and Distinguished Researcher in National ICT Australia. His awards
include the IEEE Control Systems Award of 1997, the 2001 IEEE James
H Mulligan, Jr Education Medal, and the Bode Prize of the IEEE Control
System Society in 1992, as well as several IEEE and other best paper
prizes. He is a Fellow of the Australian Academy of Science, the Australian
Academy of Technological Sciences and Engineering, the Royal Society,
and a foreign associate of the US National Academy of Engineering.
He holds honorary doctorates from a number of universities, including
Universit\'e Catholique de Louvain, Belgium, and ETH, Z\"urich. He is
a past president of the International Federation of Automatic Control
and the Australian Academy of Science. His current research interests
are in distributed control, sensor networks and econometric modelling. \end{IEEEbiography}

\end{document}